\documentclass[manuscript]{acmart}

\AtBeginDocument{%
  \providecommand\BibTeX{{%
    \normalfont B\kern-0.5em{\scshape i\kern-0.25em b}\kern-0.8em\TeX}}}

\copyrightyear{2025}
\acmYear{2025}
\setcopyright{cc}
\setcctype{by-nc-nd}
\acmConference[IUI '25]{30th International Conference on Intelligent User Interfaces}{March 24--27, 2025}{Cagliari, Italy}
\acmBooktitle{30th International Conference on Intelligent User Interfaces (IUI '25), March 24--27, 2025, Cagliari, Italy}
\acmDOI{10.1145/3708359.3712089}
\acmISBN{979-8-4007-1306-4/25/03}

\usepackage{listings}
\usepackage{soul}
\usepackage{enumitem}
\usepackage{subcaption}

\begin{document}

\title{Controlling AI Agent Participation in Group Conversations: A Human-Centered Approach}

\author{Stephanie Houde}
\orcid{0000-0002-0246-2183}
\affiliation{%
  \institution{IBM Research}
  \city{Cambridge}
  \state{MA}
  \country{USA}
}
\email{stephanie.houde@ibm.com}

\author{Kristina Brimijoin}
\orcid{0000-0001-5616-9567}
\affiliation{%
  \institution{IBM Research}
  \city{Yorktown Heights}
  \state{NY}
  \country{USA}
}
\email{kbrimij@us.ibm.com}

\author{Michael Muller}
\orcid{0000-0001-7860-163X}
\affiliation{%
    \institution{IBM Research}
    \city{Cambridge}
    \state{MA}
    \country{USA}
}
\email{michael_muller@us.ibm.com}

\author{Steven I. Ross}
\orcid{0000-0002-2533-9946}
\affiliation{
    \institution{IBM Research}
    \city{Cambridge}
    \state{MA}
    \country{USA}
}
\email{steven_ross@us.ibm.com}

\author{Dario Andres Silva Moran}
\orcid{0000-0002-3049-3139}
\affiliation{%
  \institution{IBM Research}
  \city{La Plata}
  \state{BA}
  \country{Argentina}
}
\email{dario.silva@ibm.com}

\author{Gabriel Enrique Gonzalez}
\orcid{0009-0001-4818-1205}
\affiliation{%
  \institution{IBM Research}
  \city{Necochea}
  \state{BA}
  \country{Argentina}
}
\email{gabriel.gonzalez@ibm.com}

\author{Siya Kunde}
\orcid{0000-0002-0138-3862}
\affiliation{%
  \institution{IBM Research}
  \city{Yorktown Heights}
  \state{NY}
  \country{USA}
}
\email{skunde@ibm.com}

\author{Morgan A. Foreman}
\orcid{0000-0002-2739-5853}
\affiliation{%
  \institution{IBM Research}
  \city{Houston}
  \state{TX}
  \country{USA}
}
\email{Morgan.Foreman@ibm.com}

\author{Justin D. Weisz}
\orcid{0000-0003-2228-2398}
\affiliation{
    \institution{IBM Research}
    \city{Yorktown Heights}
    \state{NY}
    \country{USA}
}
\email{jweisz@us.ibm.com}

\renewcommand{\shortauthors}{Houde et al.}

\begin{abstract}
    Conversational AI agents are commonly applied within single-user, turn-taking scenarios. The interaction mechanics of these scenarios are trivial: when the user enters a message, the AI agent produces a response. However, the interaction dynamics are more complex within group settings. How should an agent behave in these settings? We report on two experiments aimed at uncovering users' experiences of an AI agent's participation within a group, in the context of group ideation (brainstorming). In the first study, participants benefited from and preferred having the AI agent in the group, but participants disliked when the agent seemed to dominate the conversation and they desired various controls over its interactive behaviors. In the second study, we created functional controls over the agent's behavior, operable by group members, to validate their utility and probe for additional requirements. Integrating our findings across both studies, we developed a taxonomy of controls for when, what, and where a conversational AI agent in a group should respond, who can control its behavior, and how those controls are specified and implemented. Our taxonomy is intended to aid AI creators to think through important considerations in the design of mixed-initiative conversational agents.
\end{abstract}




\maketitle

\section{Introduction}

The widespread availability of conversational agents, particularly those based on large language models (LLMs), has affected how people complete tasks. It is increasingly common for people to ask an agent for assistance with a writing task~\cite{lee2024design}, a coding task~\cite{nguyen2024beginning, ross2023programmer, yen2024search}, or a summarization task~\cite{abburi2023generative, jin2024comprehensive}. Often, these tasks are implemented through conversational user interfaces in which the interaction mechanics are trivial: the (human) user makes a statement or request to the AI agent and the AI agent generates a reply via LLM inference using the conversational history as context.

In contrast to the one-on-one nature of these human-AI interactions, a large number of professional work projects occur in groups, where asynchronous and real-time group discussions take place~\cite{cutler2021meeting}. AI agent participation in such discussions could help provide additional ideas and resources that enrich both the work processes and outcomes~\cite{vats2024survey, seeber2020machines}. However, in group conversations, it is not straightforward to determine when and how an AI agent should contribute. For example, it might be inappropriate for the agent to respond to every utterance made by every member of the group in a lively discussion. Group interaction by an agent requires a more flexible strategy involving judgement of what content should be shared when.

To understand the challenges of designing an agent that ``fits in'' to a group setting, we built an LLM-based conversational agent prototype called ``Koala.'' Koala is an AI-based group discussion participant, situated in Slack\footnote{Slack. \url{https://slack.com}} and implemented as a bot application. We conducted two user studies to understand Koala's impact on the dynamics of an ideation activity and to explore the design space for its interactive behaviors. 

In our first user study, we established an initial understanding of the value of including an AI agent within a group brainstorming session as perceived by study participants, along with human-centered requirements for the agent's behavior. We engaged small groups in real-time brainstorming conversations (akin to \cite{bouschery2024artificial, shaer2024ai, suh2021ai}), in Slack channels configured with different levels of AI participation: (1) without Koala, (2) with a reactive variant of Koala that only responded when directly addressed, and (3) with a proactive variant of Koala that made its own determinations of when to contribute. Participants preferred completing the exercise with, rather than without, Koala, and they reacted positively to the ideas it generated and the way it could help keep the conversation moving. However, participants raised significant concerns about feeling overwhelmed and interrupted by the proactive variant of Koala and expressed numerous ways in which they desired to control its behavior.

Based on feedback from the first study, we created a new variant called Koala II by updating its underlying LLM, prompt, and algorithms to improve its proactive behaviors. We also created a settings panel to control both the content and display of its contributions. In a second study, we re-recruited our participants to experience Koala II. We also developed a series of design probes to explore other ways of controlling Koala's interactive behaviors than the ones we were able to implement in our settings panel. Our refinements of Koala were effective, with participants noting its improved timeliness and adherence to the topics. Participants found the ability to control Koala II's interactive behavior very useful, but they also wished to control other aspects of its behavior. These insights revealed a range of controllable aspects, and ways to control them, that culminated in the creation of a taxonomy for understanding important considerations in the design of proactive conversational agents.

Our paper makes the following contributions:

\begin{itemize}
    \item The results of a mixed-methods analysis that found that a conversational AI agent was valued as a member of a group brainstorming team, even while its interactive behaviors did not always suit the group's dynamic.
    \item A taxonomy derived from two user studies that identifies aspects of an AI agent's interactive behaviors in a group conversation along with ways those aspects can be controlled (Figure \ref{fig:control-taxonomy}).
    \item A conceptual extension of mixed-initiative interaction frameworks (e.g., \cite{fitts1951human, parasuraman2000model, sheridan1988task, shneiderman2022human, muller2022frameworks}) that challenges assumptions that an AI agent's interactive behaviors are fixed or solely determined by the workflow in which it is embedded; rather, users may dynamically adjust these behaviors throughout the course of their interaction with it.
\end{itemize}

\section{Related work}

We outline three areas relevant to our study of AI agent participation in group settings: research on mixed-initiative interfaces and frameworks for human-AI interaction, the act of brainstorming as a collaborative activity, and guidelines for the design of conversational agents.

\subsection{Mixed-initiative interfaces and frameworks for human-AI interaction}
\label{sec:mixed-initiative-interfaces} 

One of the earliest technical accounts of joint work between human and computing technologies was what has become known as the Fitts Allocation, published in 1951~\cite{fitts1951human}. Fitts proposed that certain types of work should be assigned to \emph{either} a human \emph{or} a machine, according to a set of rules about which party was best suited to perform which tasks. Newer frameworks were developed by \citet{sheridan1988task} and \citet{parasuraman2000model}, but these contained the same basic assumption that human initiative and agency traded off with technological initiative and agency.

Recent research on mixed-initiative interactions~\cite{horvitz1999principles} and mixed-initiative creative interfaces~\cite{deterding2017mixed, spoto2017mici}, along with modern human-AI interaction frameworks~\cite{muller2022frameworks, shneiderman2022human} suggest that initiative between humans and intelligent systems does not lie on a single trade-off axis; rather, they comprise a two-dimensional space in which both parties can independently act with lower or higher levels of initiative. \citet{muller2022frameworks} further demonstrate how the initiative of each actor may \emph{dynamically change} during the course of the interaction.

One limitation of these frameworks is that, although they can describe \emph{that} an AI agent can act with lower or higher levels of initiative, they do not help designers and practitioners determine \emph{how} that agent ought to behave in any given situation. Recent work has examined the design space of proactive interactions, such as by passively recommending visualizations in a data science context~\cite{tabalba2023investigation} or by determining when to interrupt a user with a notification or informational message~\cite{meck2023may, reicherts2021may}. As \citet{kraus2023improving} note, proactivity can be a ``double-edged sword'' that must be configured properly to maintain users' preferences and trust \cite{kraus2021role, meurisch2020exploring}. Without such configuration, users may resist agent-provided information, such as help services \cite{goutier2024proactive}. \citet{jain2023co} used participatory methods to create design patterns for a proactive auto-response messaging agent. \citet{zargham2022understanding} used a different approach -- visual storyboards -- to learn users' preferred patterns for what they called the ``proactivity dilemma'' in design. \citet{gammelgaard2024designing} built prototypes to compare users' responses to intermittent vs. continuous vs. proactive agent behaviors in a music recommendation application. \citet{liao2023proactive-b} summarized portions of these learnings in a full-day SIGIR tutorial.

However, such work neglects an important part of the design space: how should AI agents behave in \emph{group settings} where the interaction dynamics are more complex? With partial success, neural networks have been trained to handle the semantics~\cite{gu2022hetermpc}, relevance~\cite{wang2020response} and addressees~\cite{ouchi2016addressee} of AI responses in multi-party conversations (e.g., ``\textit{who} says \textit{what} to \textit{whom}?''). This work has not yet considered contextual factors, such as \textit{when} to interject a question or response, and how that \emph{when} depends on the task at hand~\cite{abbasian2024foundation}. For example, users of \citeauthor{oh2018lead}'s DuetDraw, an interface that allows users to draw collaboratively with AI, wanted to lead the interaction and only wanted AI explanations when requested~\cite{oh2018lead}. By contrast, \citet{koch2019may} found that 14 of 16 professional designers preferred to work with an AI that took initiative in proposing inspirational materials for creating a digital mood board \cite{koch2019may}.

\citet{mccomb2023focus} examined the use of AI within engineering design and drew a distinction between \emph{reactive} AI applications, in which AI acts in response to human activities, and \emph{proactive} AI applications, in which AI acts more as a partner or guide. Similarly, \citet{moruzzi2024user} identify a dimension of \emph{engagement} in which an AI's approach to interaction may either be ``suggesting'' or ``taking initiative.'' 

In this paper, we build upon these concepts by using a human-centered approach to explore what it means for an AI agent to act ``proactively,'' within the complexity of group conversational settings. Our work extends existing human-AI interaction frameworks by developing a new taxonomy for how agents should participate in groups.

\subsection{Group ideation as a testbed for examining AI's interactive behaviors}
\label{sec:brainstorming-1}

To explore an AI agent's interactive behaviors in group conversational settings, we needed a purpose or reason for the agent to interact with other people. We chose group ideation -- the process by which a group comes together to brainstorm ideas on a topic or theme~\cite{osborn1953applied, paulus2003enhancing} -- as it is a familiar practice, it can be conducted over a textual communications channel~\cite{wang2010groups, kerr2009beyond}, and LLM-based conversational AI agents have been shown to contribute meaningfully in this task~\cite{memmert2023towards, nomura2024towards, moruzzi2024user, shaer2024ai, muller2024group}. 

In \citeauthor{osborn1953applied}'s classic analysis of brainstorming, participants may propose ideas in any order during a \textit{divergence} phase, and participants may evaluate and select or curate a subset of those ideas for future use during a \textit{convergence} phase~\cite{osborn1953applied}. For our purposes in examining an AI agent's interactive behaviors, both phases -- divergence and convergence -- include moments in which the next conversational turn may be taken by any of the parties involved in the activity (i.e. human or AI). Brainstorming thereby presents exactly the types of interaction problems we want to investigate: \emph{when} (and implicitly, \emph{why}) should the AI contribute to the on-going discussion?

Incorporating AI is the most recent technological enhancement to group ideation, although the impact of AI agents on dyadic and group co-creativity are mixed. \citet{geerts2021brainstorming} conducted brainstorming sessions in which a scripted dialog was presented as both a wizard-of-oz AI and as a human, and found no difference in the experiences of the participants \cite{geerts2021brainstorming}. \citet{maier2022comparing} facilitated co-located groups in brainstorming sessions with human and AI facilitators, both of whom had issues such as blocking one another's contributions and making interruptions (human) and deviating from the topic and exhibiting periods of silence (AI), leaving participants frustrated. \citet{memmert2023towards} found an increase in cognitive stimulation from exposure to other team members' perspectives, but also notes the risks of free riding or reduced human effort with AI-generated contributions~\cite{memmert2023towards}. 

In contrast, both \citet{wieland2022electronic} and \citet{yu2023investigating} found that brainstorming with an AI partner, instead of a human one, was associated with an increase in the number and diversity of ideas. \citet{bouschery2023ai} evaluated nominal groups (each person brainstorms alone), interactive (human-only) groups, hybrid (human-AI) groups and AI-only groups and concluded, ``hybrid groups outperform both interactive and nominal groups in terms of the number of generated ideas and perform on par in terms of brainstorming creativity while requiring only half the human resources''~\cite{bouschery2023ai}. Within hybrid human-AI brainstorming groups, \citet{muller2024group} showed that hybrid ideas -- those that were either created or modified by at least one person and the AI -- were more likely to be identified as being the best ones generated by the group. Using a related group task (brainwriting), \citet{shaer2024ai} reported that the AI sometimes produced so much content as to interfere with the work of the people in each group. Thus, we observe that group ideation is a suitable task for our work as the interaction dynamics of an AI agent are non-trivial.

\subsection{Design guidelines for conversational agents}

Conversational AI agents have experienced widespread adoption, fueled by early advances in intent-based classification systems~\cite{freed2021conversational, hussain2019survey}, and more recently, LLM-based approaches~\cite{zhang2023instruction, ouyang2022training, liu2023summary}. Recently, a new practice of conversational UX design has emerged to focus on how to design the interactions that a human user has with conversational agents~\cite{moore2020design, moore2018conversational, moore2019conversational, silva2024towards}. These design systems focus on identifying ``recognizable interaction types... [which] frees designers from having to reinvent how users interact with the application and frees users from having to learn new interaction methods so both can focus on the content of the particular application.''~\cite[p.1]{moore2020design}. Although these design systems and sets of guidelines are useful aids to the designers of conversational AI systems, they typically focus on single-user, task-based use cases. Our work seeks to expand the design space of conversational agents by enumerating controllable elements of participation within group settings.

\section{Koala: A conversational agent for group ideation}
\label{sec:koala-I}

\begin{figure*}[ht]
    \centering
    \includegraphics[width=\linewidth]{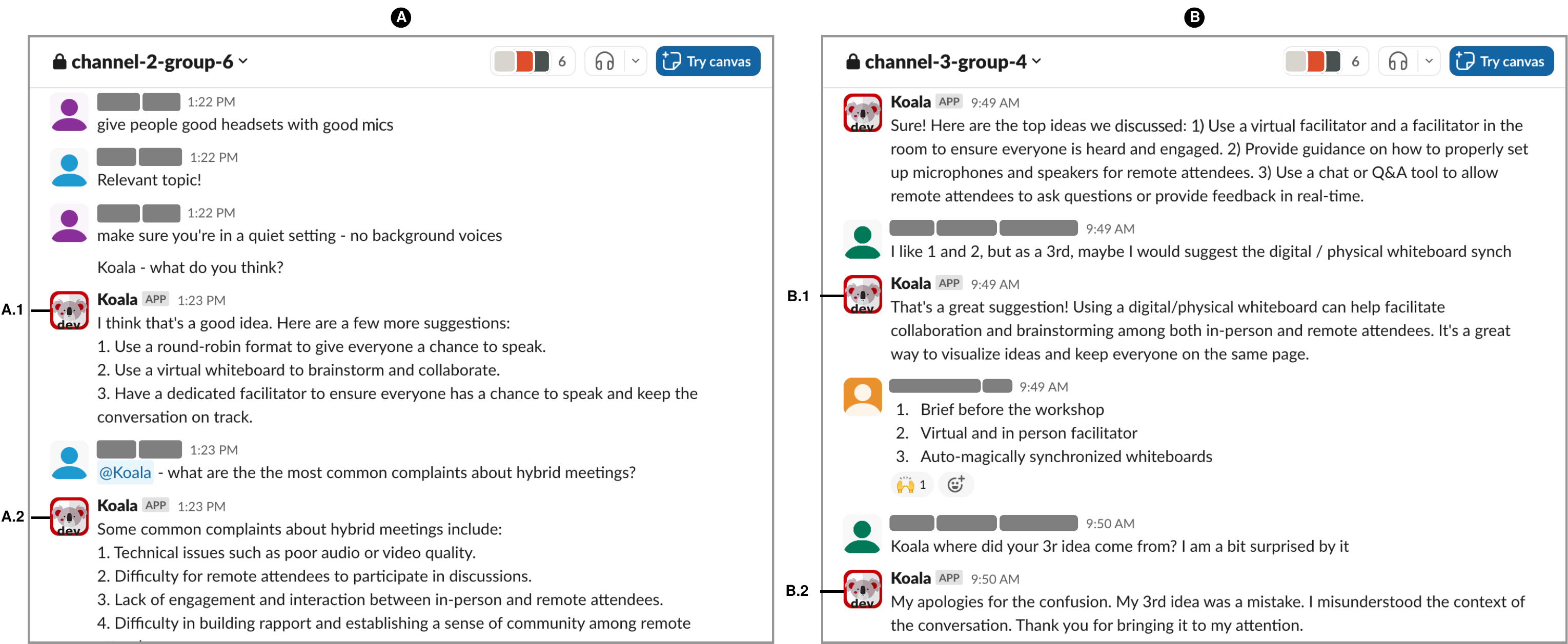}
    \caption{\textbf{Koala as an AI participant in Slack}. These screenshots (with human participant names redacted) show examples of Koala participating in a Slack channel. (A) The reactive variant of Koala replies to a user's question addressed to ``Koala'' (A.1) or to ``@Koala'' (A.2). (B) The proactive variant of Koala generates a proactive reply (B.1) to the conversation and a reactive reply (B.2) in response to a direct request.} 
    \Description{Two pictures of a Slack channel discussions that include multiple human and 1 AI participant - an AI agent called Koala. In the image on the left, Koala is shown responding to a question. In the image on the right, Koala is shown contributing the the conversation without being asked in a proactive manner.}
    \label{fig:KoalaConversation}
\end{figure*} 

To support our research goals, we developed Koala, a multi-party conversational agent designed to collaborate with people in group ideation tasks. We built Koala as a Slack application~\cite{slackAPI} to enable it to participate in group chat spaces; we specifically chose Slack because it is the primary chat application used within our organization and its API enabled us to develop an AI agent that could participate in group conversations. From a user perspective, ``Koala'' was just another participant within a Slack channel~\cite{laitinen2021slacking, seering2019beyond}, with an ``\footnotesize{APP}\normalsize'' tag to indicate that it was not a person~\cite{shneiderman2023on}.

We used the Llama 2 model~\cite{touvron2023llama} to drive Koala's conversational capability as it was a state-of-the-art model\footnote{Given the rapid pace at which AI has been advancing, the title of ``state of the art'' is a short-lived one; in Section~\ref{sec:koala-II}, we describe how the second-generation of Koala was built with Llama 3~\cite{dubey2024llama}.} that provided the highest-quality responses in our internal testing.

An important design dimension of Koala is determining when it should make a conversational utterance. Other Slack bots, such as Anthropic's Claude\footnote{Claude. \url{https://www.anthropic.com/index/claude-now-in-slack}}, require users to explicitly mention the bot's name to elicit a response from it (i.e. ``@''-ing the bot). Given our desire to experiment with proactive ways of participating, we developed two configurable variants of Koala: a \emph{reactive} variant that only responds when directly addressed (by either ``@Koala'' or ``Koala''), and a \emph{proactive} variant that responds \emph{when it determines that it has something valuable to say.} Koala can contribute to the conversation in the same manner as other human users, by sending textual messages as well as emoji-based reactions to existing messages. We show a screenshot of Koala in Figure~\ref{fig:KoalaConversation}.

\subsection{Determining when to speak: Koala's proactive behaviors}
\label{sec:deciding-when-to-respond}

\begin{figure*}[ht]
    \centering
    \includegraphics[width=\linewidth]{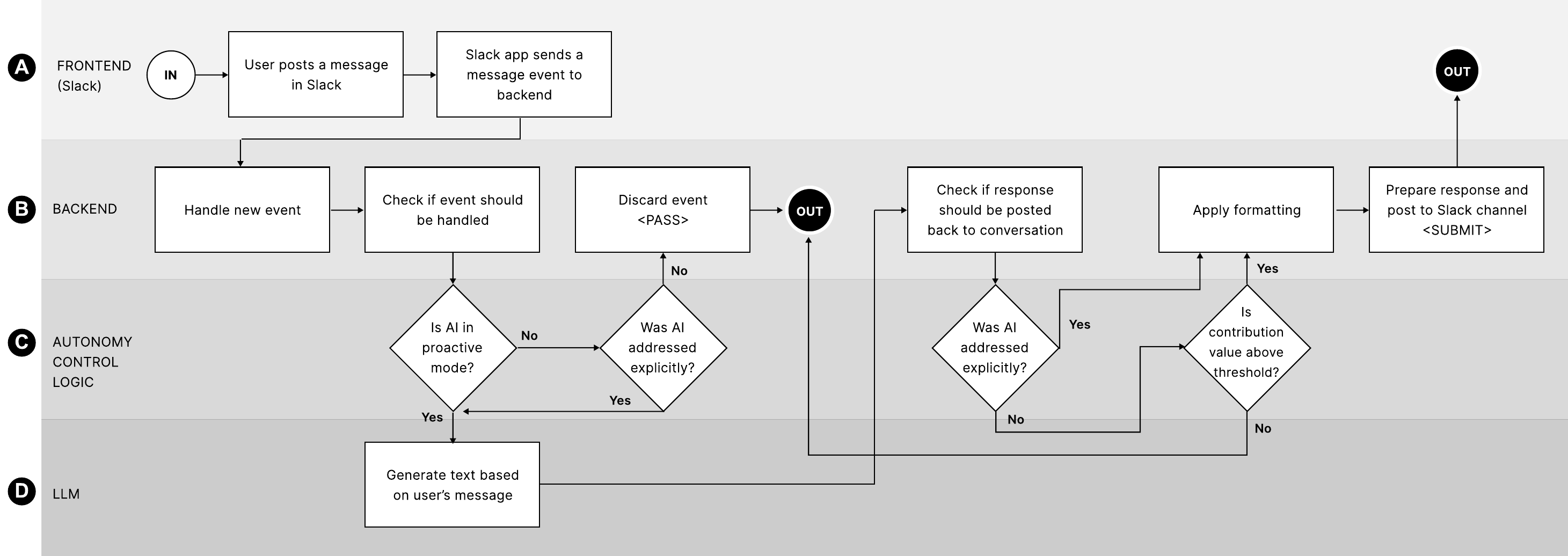}
    \caption{\textbf{Koala operational logic}. When users post a message in a Slack (A), the post triggers an event that is handled by the Koala backend (B) where control logic determines whether Koala should either immediately pass on replying or hand off for further evaluation in the autonomy control logic (C) where the LLM (D) generates a response that is further evaluated for potential posting in the channel.}
    \Description{A flow diagram of the Koala contribution workflow}
    \label{fig:koala-operation}
\end{figure*}

Conversational AI agents typically engage in one-on-one conversations with human users, where each message from the user receives a response from the agent. In a group chat where multiple people are talking to (and sometimes on top of) one another, the decision of when a proactive agent should respond to any particular user message becomes non-trivial; some mechanism is needed to determine when the agent should respond to a user's message~\cite{ganesh2023survey, gu2022hetermpc, ouchi2016addressee}.

In developing the proactive variant of Koala, we knew that writing imperative decision logic to determine what constitutes a valuable contribution would prove impossible; therefore, we relied on the underlying LLM to not only produce conversational responses in response to users' messages, but also to score those responses for the extent to which they made a valuable contribution to the conversation. We outline this process in Figure~\ref{fig:koala-operation}; at a high level, Koala is instructed to not reply to messages where it is clearly not the intended recipient, and to reply to messages when it estimates that the response it generated makes a valuable contribution to the conversation. We note that our proactive variant of Koala represents a \emph{limited form} of proactivity as it's process is triggered by the receipt of a message from a human user; a truly proactive variant of Koala would also be able to send messages to the group even when the human participants remain silent \cite{yorke2012proactive}. We provide a complete listing of the prompts used by Koala to generate responses to members of the group and score them for their value in Appendix~\ref{appendix:koala-prompt}.

During Koala's development, we noticed that the underlying LLM was unreliable in identifying the intended target for a chat utterance. Thus, we incorporated additional, external control logic to force Koala to reply if it's name was detected in a user's message, and suppress Koala from replying under other circumstances, such as when the names of one of the other participants were mentioned, regardless of how the LLM scored the value of the reply's content.

\subsection{User studies \& ethics statement} 

We conducted two user studies with Koala to understand the impact of its proactive behaviors on group ideation (Section~\ref{sec:koala-study-1}) and develop a robust understanding of the ways in which users would like to control those behaviors (Section~\ref{sec:koala-study-2}). Our research was conducted in accordance with our institution's internal policies on human-participant research. Specifically, participants provided their informed consent, participation was voluntary and could be withdrawn at any time, and all collected data were anonymized before analysis and publication.

\section{Study 1: Group brainstorming with Koala}
\label{sec:koala-study-1}

Our first study focused on understanding people's experience working with the reactive and proactive variants of Koala in a series of group ideation scenarios. We designed this study to address two research questions:

\begin{itemize}
     \item \textbf{RQ1.1}. How does the participation of an AI agent transform the brainstorming process?
     \item \textbf{RQ1.2}. How do people characterize the experience of brainstorming with a reactive vs. proactive AI agent?
\end{itemize}

\subsection{Method}

We recruited small groups of IBM employees to participate in a series of virtual brainstorming tasks with (or without) Koala as an active participant. Groups performed three brainstorming tasks, with each task focused on a different variant of Koala's participation:

\begin{itemize}
    \item \textbf{No AI}, in which Koala was not present in the channel. This condition enabled us to establish a baseline user experience of brainstorming on Slack.
    \item \textbf{Reactive AI}, in which Koala acted as another participant in the brainstorm, but was configured to only respond when directly addressed.
    \item \textbf{Proactive AI}, in which Koala actively participated in the brainstorm and made its own decisions about when to contribute. As with the reactive variant, the proactive variant always responded when directly addressed.
\end{itemize}

We used two tools to coordinate the study and run the brainstorming activities: participants joined a web conference to provide their informed consent and receive instructions for the task, then they muted themselves and turned their attention to Slack where the brainstorming activities occurred. Each session was facilitated by a human moderator, who introduced the brainstorming activity, clarified any confusions about the topic, and maintained time constraints. At no point did the moderator influence brainstorming outcomes.

\subsubsection{Brainstorming task and topics}
\label{sec:brainstorming-topics}

Groups were provided with three minutes to brainstorm on a topic with (or without) Koala. Pre-testing indicated that this short amount of time was suitable for producing a meaningful number of ideas without the session feeling protracted. We developed three topics that were relevant to a generic work environment and non-specific to any particular job role or expertise.

\begin{itemize}
  \item \textbf{Topic A}. How can we improve hybrid meetings for remote participants?
  \item \textbf{Topic B}. How can we better engage employees to use a chatbot for HR services?
  \item \textbf{Topic C}. What kinds of giveaways should we take to a customer conference?
\end{itemize}

We assigned topics in a counterbalanced fashion across the three conditions to avoid an order effect. However, each group experienced the three conditions in the \emph{same order} (i.e. first No AI, then Reactive AI, then Proactive AI), as we wanted to probe participants on their experiences with increasingly autonomous levels of AI involvement.

\begin{figure*}[ht]
    \centering
    \includegraphics[width=\linewidth]{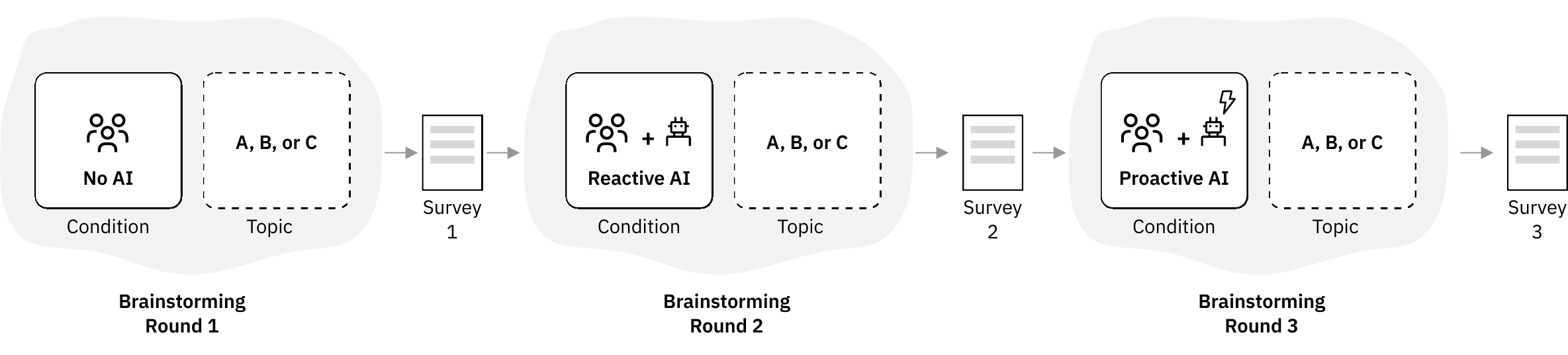}
    \caption{\textbf{Study 1 overview}. During each session, a group of participants sequentially completed three rounds of brainstorming and post-brainstorm surveys. The order of conditions (No AI, Reactive AI, Proactive AI) was kept the same for each group to assess the impact of increasing levels of AI autonomy on participants' experience. Brainstorming topics were assigned in a counterbalanced fashion to negate order effects.}
    \Description{Graphic showing the organization of each study session.}
    \label{fig:study-overview}
\end{figure*}

At the conclusion of the three minutes, we asked participants (including Koala) to select the three top ideas that they felt were the strongest of the ones they ideated, although the human participants had final decision authority. Then, we had participants fill out a short survey that asked about their experience. We show a depiction of the overall structure of the study in Figure~\ref{fig:study-overview}. Participation in our study took approximately one hour, for which each participant was compensated with the equivalent of \$25 USD. 

We collected numerous data from each brainstorming session: transcripts of the brainstorming conversations (e.g. the Slack messages sent by each participant, plus Koala), observational notes taken by the study moderator, and surveys after each brainstorming topic.

\subsubsection{Participants}
\label{sec:study1-participants}

We conducted a total of six study sessions, each having three (human) participants, for a total of 18 participants. In our analysis, we refer to individual participants using the notation \texttt{Pk.m}, where \texttt{k} is the number of the group in which they participated (1-6), and \texttt{m} is their identifier within the group (1-3)\footnote{We use this same notation in Study 2 as the participants were the same people.}.

All participants were employees of IBM. Our participants included 10 who identified as female (55.5\%), 7 who identified as male (38.8\%), and 1 who preferred not to disclose their gender identity (5.5\%). Geographically, two-thirds of our participants were from the United States (12; 66.6\%), followed by Argentina (3; 16.6\%) and Germany (3; 16.6\%). Our participants' job roles included content strategy, user research, machine learning, UX design, software engineering, and video production. They reported interacting with conversational AI assistants at least once a week or more frequently (7; 38.8\%), interacting with them about once a month (9; 50\%), or interacting them less frequently (2; 11.1\%).

We recruited participants in groups of three to ensure groups had a minimum of two participants in the event a participant had to cancel. Since prior work has shown that groups of people who are familiar with each other are more productive in ideation tasks than ad-hoc groups~\cite{dennis1990adhoc, paulus2012collaborative} due to their ability to communicate more effectively~\cite{pinsonneault1999electronic}, we recruited people who had prior working relationships with each other.

\subsubsection{Survey}

We developed a post-task survey, with minor variations for each of our three conditions, that asked open-ended questions about whether participants felt the session was effective, whether they were happy with the ideas that were produced, whether their team worked well together, and which conditional experience they preferred. We also included a set of questions, based on the Creativity Support Index~\cite{dewit2020using} and thesis work of \citet{asio2015empirical} to gauge the extent to which people felt Koala contributed to their creative process. We made minor modifications to these validated scales to fit the conversational AI modality of our research. We provide a listing of all survey questions in Appendix~\ref{appendix:study-1-survey}.

\subsection{Results}

\subsubsection{RQ1.1. Koala's impact on the brainstorming process}

\begin{figure*}[t]
    \centering
    \begin{subfigure}{0.48\textwidth}
        \includegraphics[width=\textwidth]{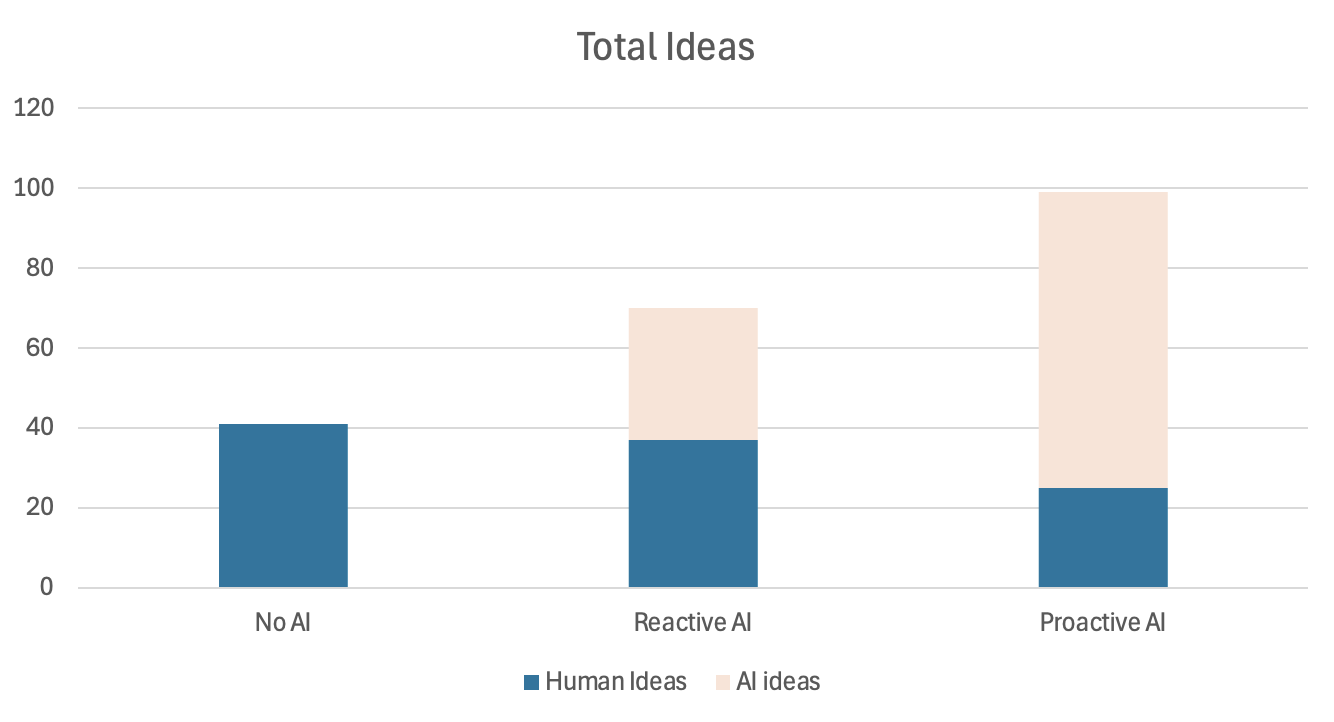}
        \caption{}
    \end{subfigure}
    \begin{subfigure}{0.45\textwidth}
        \includegraphics[width=\textwidth]{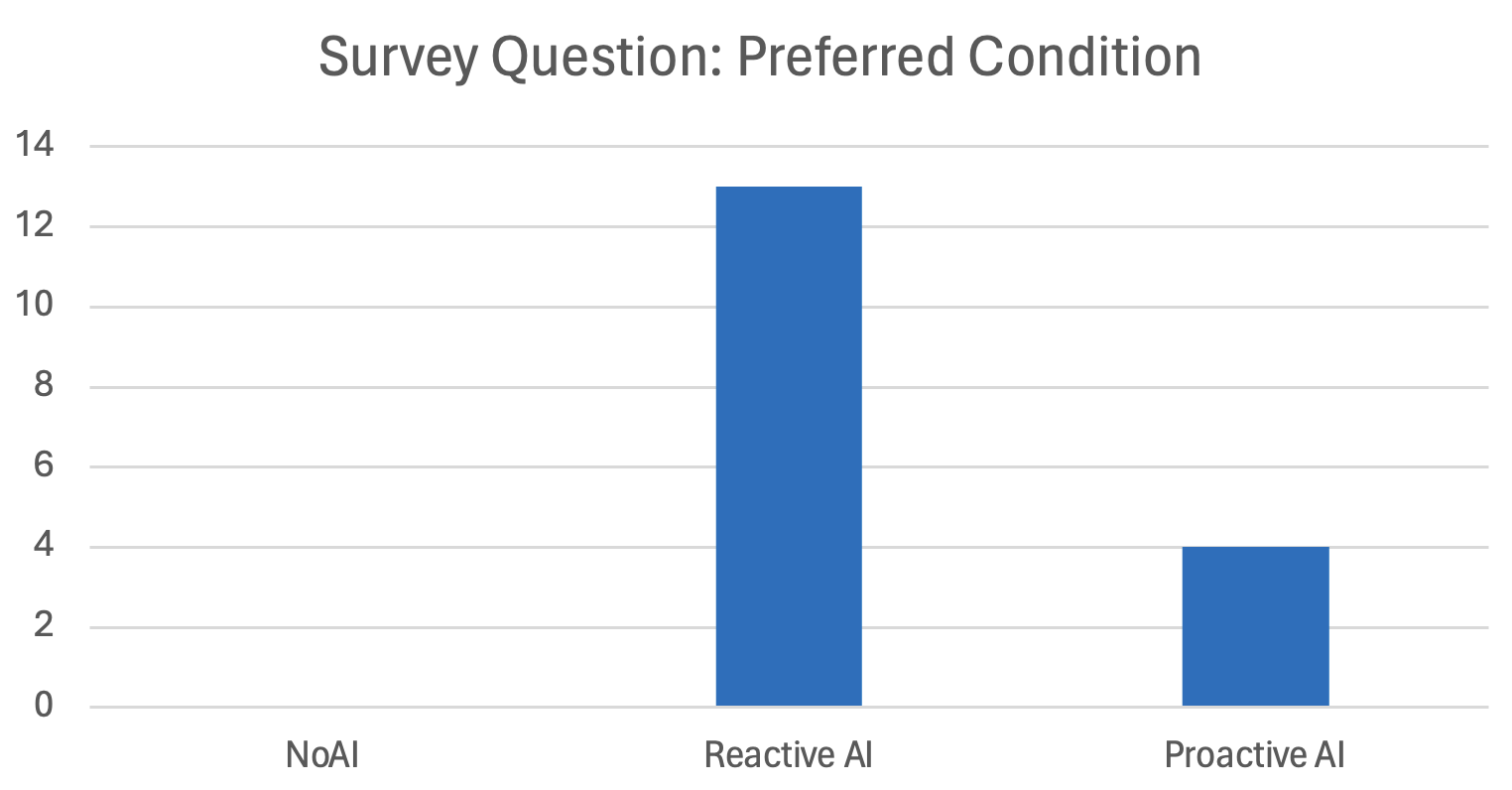}
        \caption{}
    \end{subfigure}
    \caption{(a) Impact of Koala on the production of ideas and (b) participants' preferences for the different Koala variants.}
    \Description{The figure contains two histograms. (a) On the left, there are total counts of ideas in each of the three AI-conditions. The No AI condition has the fewest ideas, followed by the Reactive AI condition, with approximately equal idea counts from humans and from AI. The Proactive AI condition has the most ideas, with about 75\% of those ideas coming from the AI. (b) On the right, there is a summary of responses to a survey question about which condition was preferred by each human. The majority of respondents preferred the Reactive AI condition, followed by a much small number of responses for the Proactive AI. Importantly, no human preferred the No AI condition.}
    \label{fig:study-1-koala-impact}
\end{figure*}

To understand how Koala impacted the brainstorming process, we conducted several analyses of our data. First, we considered the number of ideas produced within each group across the different conditions, as determined by a coding process by three authors (Figure~\ref{fig:study-1-koala-impact}a). We observed that 73\% of all ideas produced during either of the AI conditions were contributed by Koala. Not only was the volume of ideas increased with Koala's presence, but Koala's ideas also comprised a third (33\%) of all of the top ideas. Participants also preferred having either Koala variant in their group over not having it, with the reactive variant receiving the most preference (Figure~\ref{fig:study-1-koala-impact}b). These results suggest Koala had a positive impact on the outcomes of the brainstorming task\footnote{For a more extensive analysis on how brainstorming ideas evolved between participants and Koala, please refer to \citet{muller2024group}.}.

We also considered how Koala impacted participants' work process. We conducted a reflexive thematic analysis~\cite{braun2021can} on open-ended survey responses and text transcripts of the brainstorming sessions. We used an inductive approach in which three researchers collaboratively labeled these data, iterating until they came to an agreement on important themes. Through this process, we constructed three top-level themes with a number of subthemes to identify how Koala impacted the work process. The thematic analysis made it clear that Koala had a mixed effect. It provided participants with valuable support (\textbf{Advantages}), such as by summarizing ideas contributed across the group or helping get the brainstorming off the ground. But participants also talked about how Koala's participation was distracting and intrusive (\textbf{Disadvantages}), especially when it contributed too much, too often. Finally, we noted that participants desired to improve Koala's behavior (\textbf{Improving AI behavior}) via some kind of control mechanism.

\paragraph{Theme 1} \textbf{Advantages}

\begin{itemize}
    \item \textbf{Help getting started}. Almost half of participants commented that the reactive variant helped their group get started in their brainstorming task. P2.3 spoke of \textit{``removing the `white page' problem,''} and P6.1 remarked, \textit{``Koala was able to add to the conversation when we were getting stuck,''} indicating that Koala's ideas were useful throughout the brainstorming session. 

    \item \textbf{Perceived speed}. Several participants indicated that compared to not having Koala, the reactive variant of Koala sped up the process. P2.2 wrote, \textit{``It was  more fast working with Koala,''} and P2.3 said, \textit{``Koala made it much more fluid and expedited.''}

    \item \textbf{Structure}. Some participants noted that Koala provided structure to their session. In P6.2's experience, \textit{``Koala's suggestions helped structure the conversation... Koala added structure and gave us a springboard for ideas. It served as a psuedo-moderator [sic].''}

    \item \textbf{Summaries}. Nearly half of participants made positive comments about the value of Koala's ability to provide summaries of ideas. P4.1 said, \textit{``I liked the structuring/summarization function of Koala,''} and P3.1 similarly commented, \textit{``The summarizations that Koala created were valuable and helped keep us more focused.''} 

    \item \textbf{Validation}. Some participants commented that seeing ideas from Koala that were similar to their own gave them a sense of validation. P1.1 wrote that being able to ask Koala specific questions \textit{``allowed our team to be creative and come up with relevant ideas, verify them with Koala, then move forward.''}

    \item \textbf{Information}. A few participants noted that it was useful to have Koala fill in knowledge gaps as opposed to offering new ideas. P1.3 commented, \textit{``Koala was helpful at providing data on issues that I didn't know about.''}

    \item \textbf{Human-AI collaboration}. Several participants commented on the benefits of having both people and Koala in the discussion. P4.1 wrote, \textit{``The collaboration was much better through using the chat together with the Koala ideas that we could all see. For me it was interesting to see what the others asked Koala to get some inspirations for my own prompts.''} P4.2 said, \textit{``I think we worked very well together because we equally spoke to each other and to Koala in the chat.''} P2.3 similarly remarked on the value of human-AI collaboration, saying, \textit{``Very innovative ideas resulted combining innovative ideas from people to ask Koala for concrete ideas.''}
\end{itemize}

\paragraph{Theme 2} \textbf{Disadvantages}

\begin{itemize}
    \item \textbf{Disruptive proactivity}. The primary complaint about Koala was in regard to its proactive variant. Many participants felt its contributions were intrusive and caused them to feel distracted and overwhelmed. We discuss these issues in detail in Section~\ref{sec:impact-of-proactivity}, which motivated changes to Koala's behavior (Section~\ref{sec:koala-II}).

    \item \textbf{A stifling effect}. Participants indicated that in some instances, Koala's presence had a stifling effect on human conversation and creativity. P5.1 commented that, \textit{``The session started [with humans] asking Koala to answer the question, which then led to mostly narrowing down the choices Koala presented. There was less room for expressing unique ideas as a result.''} P1.2 noted, \textit{``Sometimes brainstorming or great ideas will come from out-of-box ideas but I felt I boxed myself.''} These stifling effects may be related to an online form of production blocking (e.g.,~\cite{nijstad2003production, stroebe2010beyond}).

    \item \textbf{Inaccurate answers}. Approximately one-third of participants noticed inaccuracies in Koala's responses or summaries. P2.3 observed that the summaries provided by Koala included \textit{``summaries or top themes that handn't even been mentioned in the chat.''} P3.2 simply said, \textit{``the summarization was incorrect,''} and P1.2 said that Koala \textit{``struggled especially [...] to present the ideas generated in slack.''} Such issues are common with LLMs~\cite{hicks2024chatgpt} and mitigating them is an active area of research (e.g.~\cite{tonmoy2024comprehensive}).
\end{itemize}

\paragraph{Theme 3} \textbf{Improving AI behavior}

\begin{itemize}
    \item \textbf{Regulating AI behavior}. Participants noted behaviors of the proactive variant of Koala that they found to be objectionable (Section \ref{sec:impact-of-proactivity}). These complaints were sometimes accompanied by requests for improved default behaviors about when and how much Koala should contribute. For example, P4.3 suggested that Koala should \textit{``Give some extra time to answer, so you allow real people to answer first''}. P4.2 wondered if perhaps \textit{''Koala could ask 'Would you like me to share the top 3' instead of writing a big sentence without being asked.''}
    \item \textbf{User control of AI}. Beyond wanting more regulated default behaviors, some participants attempted to actively control Koala's behavior within the chat. We discuss this point further in Section~\ref{sec:controlling-the-ai}. 
\end{itemize}

\subsubsection{RQ1.2. Reactive versus proactive behaviors}
\label{sec:impact-of-proactivity}

As previously seen in Figure~\ref{fig:study-1-koala-impact}b, the majority of participants (13; 72.2\%) preferred interacting with the reactive variant of Koala. Participants indicated a number of reasons why they preferred the reactive variant to the proactive variant.

\begin{itemize}
    \item \textbf{Distraction}. Multiple participants across all six groups commented that the proactive variant distracted their team from the task. P3.3 commented on proactive Koala, saying, \textit{``[it] was distracting and didn't help at all, and picked one of its own ideas as a favorite.''} P6.1 indicated that proactive Koala \textit{``was too talkative, both in length of message and frequency.''} 
    \item \textbf{Intrusiveness}. At least half of participants commented that the proactive variant intruded too much on their group's conversation. Koala jumped in at the wrong times, wrote too much, responded too frequently, and generally got in the way of the humans having the discussion. P2.1 remarked, \textit{``I think Koala was too much proactive at the beginning which lead us the humans to lose focus.''} P6.2 was more pointed in their feedback, saying, \textit{``Koala dominated the conversation. It felt like a pedantic student who wouldn't create space for others to participate.''} P5.2 felt that their team \emph{``...did generate less ideas though, but that maybe because they were interrupted a lot by Koala. [...] A room for improvement would be to ask Koala to pause for some time before she gets back into the conversation.''}    
    \item \textbf{Feeling overwhelmed}. Participants expressed feelings of being \textit{``overwhelmed''} (P2.3) by proactive Koala, likely in reaction to its distractions and intrusions. P4.2 explained, \textit{``I was having a hard time building up on the ideas from the others, because everything moved so fast.''} P5.1 similarly felt, \textit{``Koala was helpful, but also produced so much text so quickly that it was hard to keep up with the conversation.''}
\end{itemize}

Some participants saw potential benefits in having some degree of proactive involvement from Koala, but in its current form, the benefits of proactivity were not enough. P1.3 explained, \textit{``I liked channel 2 [Reactive AI] \& 3 [Proactive AI] best. I think i would pick channel 3 [Proactive AI] if there was a way to temporarily disable koala if needed.''} P4.1 similarly said, \textit{``I think I would prefer a combination of Channel2 [Reactive AI] + Channel3 [Proactive AI] - where Koala can add suggestions when not answered, but not always as the very first person. It felt a bit like we `lost the race against the machines,' because Koala's answers were always soo [sic] quick and mostly eloquent.''} 

\subsubsection{Desire to control interactive behaviors} 
\label{sec:controlling-the-ai}

We observed several attempts by participants to control the interactive behaviors of Koala during the brainstorming sessions. For example, P6.3 asked proactive Koala to cease its participation, saying, \textit{``koala - leave the rest to us, ok?''} and later in the same session, \textit{``Koala - thanks for these ideas. We'll ping with a question if you're needed.''} But, not all attempts at controlling Koala's behavior were aimed at reducing or minimizing its participation; sometimes, participants made comments to encourage more participation from Koala. For example, P3.1 asked reactive Koala to take initiative: \textit{``Can Koala help choose the 3 solutions?''} P1.1 and P1.2 both provided encouragement to Koala, as well as an additional opportunity to take a conversational turn, when they said (at different times), \textit{``Great start, Koala.''}

These attempts at behavioral control, coupled with comments expressing such desires (\textbf{Improving AI behavior}), led us to revisit our implementation of Koala's interactive behaviors and conduct a deeper examination of exactly what kinds of controls are needed to make Koala a more compatible group participant.

\section{Koala II: Exploring dimensions of control over interactive behaviors}
\label{sec:koala-II}

Based on participants' experiences in Study 1, we revised Koala to (1) improve its proactive behaviors (Section~\ref{sec:koala2-proactivity}), and (2) parameterize some of its behavioral attributes and expose them in the user interface for users to control (Section~\ref{sec:koala2-user-control}).

\subsection{Enhancements to proactive behaviors}
\label{sec:koala2-proactivity}

We made a number of changes to Koala's proactive variant based on participants' feedback that it was distracting, intrusive, and left them feeling overwhelmed.

\begin{itemize}
    \item We switched the underlying LLM used by Koala to Llama 3~\cite{dubey2024llama}, which was released after we ran Study 1. In our testing, we found the new model was significantly less likely to hallucinate or produce inaccurate answers, due in part to a doubling of its token context length and increased capacity to incorporate more of the conversational history in its prompts.
    \item We revised Koala's prompt to enhance the quality of its responses (detailed in Appendix~\ref{appendix:koala-prompt}). We prompted Koala to provide more targeted and collaborative suggestions, to provide constructive criticism on existing ideas, and to contribute novel ideas, all while remaining mindful of not dominating the conversation.
    \item We simplified the process by which Koala used to assess the relevance of its ideas. Previously, Koala made an overall determination -- a score from 0 to 100 that evaluated the value of its response to the group -- to decide whether a generated response should be sent to the group, based on a fixed threshold. Now, we made this threshold configurable, allowing participants to adjust it at any time during the conversation.
\end{itemize}

\subsection{Exposing behavioral controls}
\label{sec:koala2-user-control}

Participants in Study 1 wanted explicit control over Koala's interactive behaviors. Prior work on human-AI interactions with generative AI systems has explored ways to give users various controls over the outputs of the underlying generative model (e.g.~\cite{chong2021interactive, dang2022ganslider, louie2020cococo, louie2020novice}), but we recognized the types of controls users needed were not solely related to the \emph{content} of Koala's messages, but also to the \emph{ways} those messages were contributed within the group setting. We also recognized that these types of controls were ones that could not be manifested solely through prompt engineering and that external mechanisms would be needed to implement them.

\begin{figure}[t]
	\centering
	\includegraphics[width=.55\linewidth]{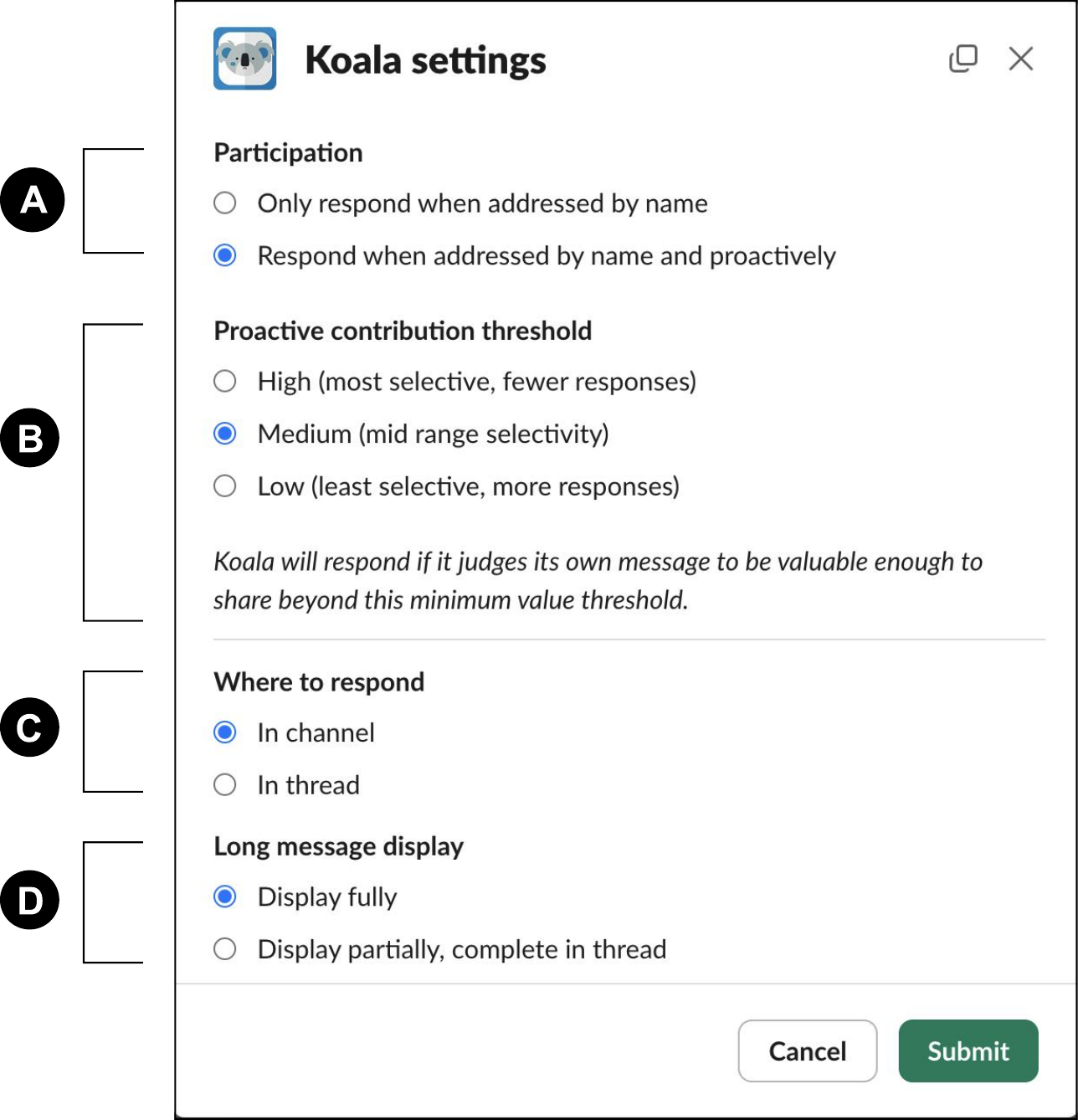}
	\caption{\textbf{Control settings}. Koala II's control settings provides users with the ability to control aspects of its interactive behaviors. These settings include (A) switching between proactive and reactive variants; (B) adjusting the threshold at which the proactive variant deems a generated response valuable to send to the group; (C) selecting whether Koala's responses should appear in the conversation or in a thread; and (D) managing how long messages are displayed.}
	\Description{Graphic showing the control settings dialog}
	\label{fig:control-settings-dialog}
\end{figure}

We built a control panel for Koala (Figure~\ref{fig:control-settings-dialog}) based on our thematic analysis and an assessment of what was technically feasible within the Slack environment. These controls were accessible at any point in time and could be changed by any member of the group.

\begin{itemize}
    \item \textbf{Proactive versus reactive toggle} (Figure~\ref{fig:control-settings-dialog}a). We added the ability for users to choose which type of Koala variant they preferred: the proactive variant that made its own decisions for when to make a conversational utterance, or the reactive variant which only replied to users when directly addressed.
    \item \textbf{Proactive contribution threshold} (Figure~\ref{fig:control-settings-dialog}b). Given our changes to how Koala scored its potential response, we exposed a coarse contribution threshold for when its response should be sent to the group. Through experimentation, we set the ``high'' threshold at 90, the ``medium'' threshold at 75, and the ``low'' threshold at 50 on the 0-100 scale used by Koala.
    \item \textbf{Where to respond} (Figure~\ref{fig:control-settings-dialog}c). Slack allows messages to be posted in two places: within the channel itself, or as a threaded reply to a message. Previously, Koala would only place its replies in the channel, but we exposed an option to have Koala reply in a thread to make it clearer to which messages Koala's replies were targeted. We added this option due to our observations that Koala would sometimes take time to reply to messages after the conversation had already moved on, making it difficult for participants to understand for which message Koala's response was intended.
    \item \textbf{Long message display} (Figure~\ref{fig:control-settings-dialog}d). Participants felt overwhelmed by Koala's responses in part because they could potentially be very long (such as when Koala offered enumerated lists of ideas rather than producing them one at a time). We added an option for those cases in which Koala produced a long comment (> 1,000 characters) to place a truncated version in the chat with the full version appearing in a threaded response.
\end{itemize}

We recognize that these controls represent only a portion of the potential design space, so we conducted a second study to gather additional feedback and synthesize all of our findings into a more robust taxonomy (Figure~\ref{fig:control-taxonomy}).

\section{Study 2: Group brainstorming revisited}
\label{sec:koala-study-2}

Participants in Study 1 felt that the proactive variant of Koala could be beneficial, so long as its behavior could be tempered to avoid the distraction, intrusiveness, and overwhelming feelings they experienced. Nine months after the first study, we ran a second study to evaluate whether our changes to Koala's behaviors, along with the addition of behavioral controls, made a difference in the quality of the experience. In addition, we aimed to elicit additional types of controls needed for conversational AI agents such as Koala using both the control settings panel (Figure~\ref{fig:control-settings-dialog}) and additional design mockups (Figure~\ref{fig:control-UI-options}) as design probes. Thus, we designed our second study to address the following research questions:

\begin{itemize}
    \item \textbf{RQ2.1}. How did Koala II's modified proactive behaviors impact peoples' experiences?
    \item \textbf{RQ2.2}. What additional types of controls are needed over an AI agent's interactive behaviors?
\end{itemize}

\subsection{Method}

We used the same method for Study 2 as Study 1, except we reduced the number of brainstorming tasks to two instead of three to make time for a discussion on controlling an AI's interactive behaviors.

\subsubsection{Participants}

To make comparisons with Koala I, we re-recruited the 18 people who participated in Study 1. Of these, a total of 14 were able to participate in Study 2. The composition of the groups remained the same as in Study 1, except that for four out of the six original groups, one of the original participants was unavailable. As before, each session took approximately 1 hour and each participant was compensated with the equivalent of \$25 USD. 

\subsubsection{Brainstorming task and topics}

Groups were provided with three minutes to brainstorm on a topic, then decide upon three top ideas. We used a new topic for the first brainstorming session:

\begin{itemize}
    \item How can we foster a sense of community and belonging within a team?
\end{itemize}

Following the session, participants engaged on the video conference in a semi-structured discussion with the moderator on the following topics:

\begin{enumerate}
    \item Compare the experience of Koala II with Koala I, to the best that they could recall,
    \item Discuss anything they wished to be different about their experience in brainstorming with Koala II, and
    \item Probe on ways they would like to control Koala II's interactive behaviors.
\end{enumerate}

Next, participants were introduced to Koala II's control settings dialog (Figure~\ref{fig:control-settings-dialog}). The moderator continued the semi-structured discussion by explaining the intent behind each control option, probing whether the given controls were useful, and eliciting further ideas on additional controls that might be missing. Participants were then asked to discuss whether and how they would like to change any of the settings for their second brainstorm.

They then conducted a second brainstorming session, on another new topic:

\begin{itemize}
    \item What are some ways to recognize and reward team members for their contributions?
\end{itemize}

At the end of the second brainstorm, the moderator asked each participant to:

\begin{enumerate}
    \item Provide a rating for how useful the controls were, on a scale of 1 (not useful) to 5 (very useful) and explain their reasoning.
    \item Discuss their feelings on who should have permission to alter the controls during a session, as the controls modify AI behavior experienced by the entire group.
\end{enumerate}

\begin{figure*}[t]
    \centering
    \includegraphics[width=.9\linewidth]{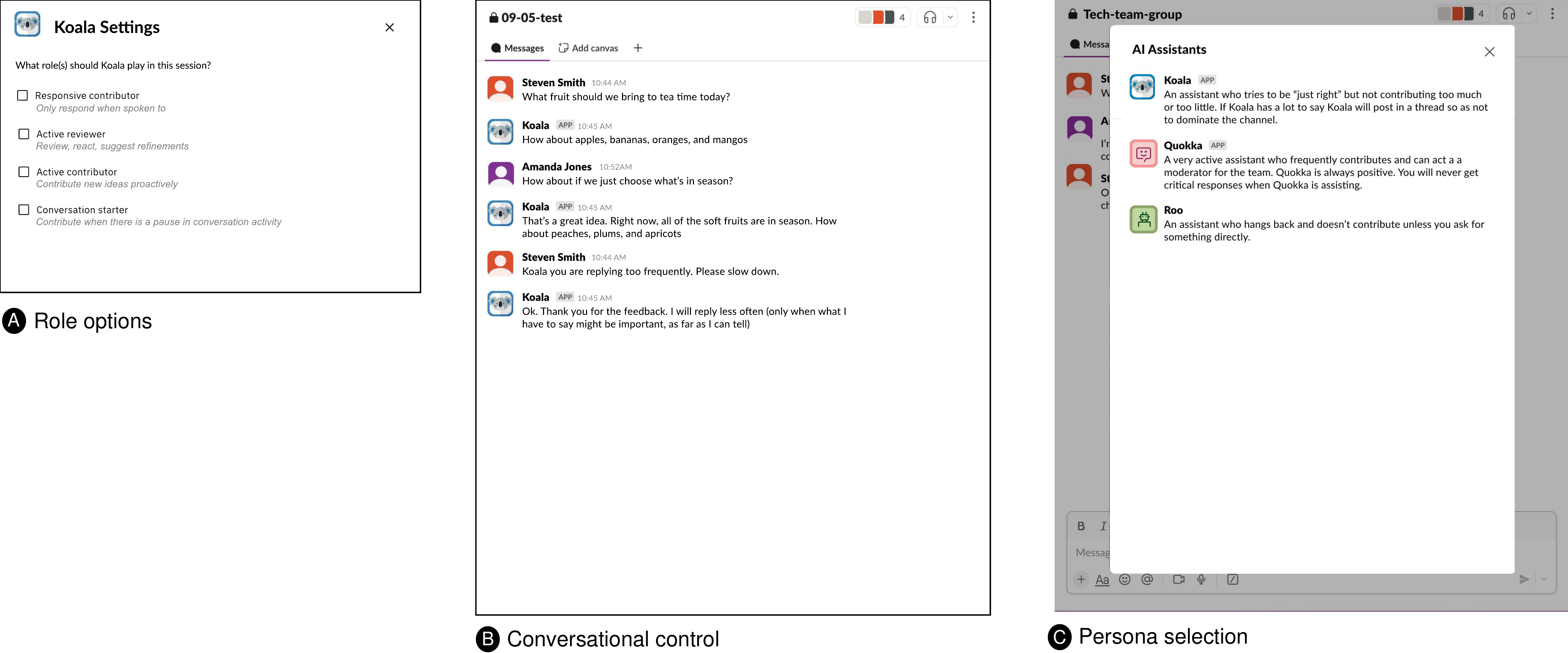}
    \caption{\textbf{Alternate agent control design mockups}. Mockups presented to participants in Study 2 show three alternative designs (with fictional users) for controlling Koala II's behavior: (A) Role options allows users to assign the role that Koala II should play in the session; (B) Conversational control enables users to adjust Koala II's settings through natural language in the chat; and (C) Persona selection allows users to select different assistant personas to fit their interaction needs.}
    \Description{Graphic showing mockups presented to users.}
    \label{fig:control-UI-options}
\end{figure*}

Finally, in the last phase of the study, the moderator presented participants with three mockup interfaces showing additional ways Koala II's interactive behaviors might be controlled. These options were motivated by observations and comments made by participants in Study 1: by selecting a higher-level role within the group (Figure~\ref{fig:control-UI-options}a), within the conversation itself through natural-language requests (Figure~\ref{fig:control-UI-options}b), or by choosing a persona with specified behavioral attributes (Figure~\ref{fig:control-UI-options}c). After a short discussion on these different options, participants were asked to rank-order their preferences.

\subsection{Results}

\subsubsection{RQ2.1. Reactions to Koala II's proactive behaviors}

Participants perceived an improvement in Koala II's proactive behaviors compared to Koala I. Koala II was observed to be quieter and less invasive during conversations, and consequently, less proactive in its participation. P2.1 reported, \emph{``The first time we used it, it was very proactive. Now I noticed it was more quiet. It reacted at the right pace. I felt it better than before.''} P4.2 recalled \emph{``feeling interrupted in the last time and had to spend a lot of time reading and not writing; [the interaction with Koala was] a lot better today.''}

Interactions with Koala II were also reported to be more comfortable and natural. P2.3 stated, \emph{``I felt much more comfortable with this version [...] Koala wrote at the perfect time, when I was about to ask, it was giving some suggestions.''} Similarly, P5.3 mentioned, \emph{``I felt it was a very natural interaction, more so than the last time.''} Additionally, participants noted that Koala II stayed more on topic. P6.2 mentioned, \emph{``Koala seemed to follow the same thread as the team, whereas last time it went off in different directions.''} P1.3 said, \emph{``I do remember being a little bit annoyed last time [...] I was trying to type out a thought and [Koala] was taking the conversation in different direction by writing a bunch of things unprompted. [...] That was a little less frequent this time. It was a little bit more on topic.''}

Although participants felt the new proactive behaviors were an improvement, they also pointed out other issues with Koala II's behavior. Some participants felt Koala II was \emph{``over the top''} (P3.3) and \emph{``irritating''} (P3.2, P5.1, P1.3). P5.1 reacted to Koala II's overly-positive tone: \emph{``I noticed Koala said a few times `that's a great idea!'. It was very acting like a person, and it was a little off-putting to me. I would rather it be a bot and just kind of be more neutral in its expressions.''} Hence, we conclude that Koala II did provide an improved user experience, though further improvements are still possible.

\subsubsection{RQ2.2. Controlling an AI agent's interactive behaviors}

Participants were shown the control settings dialog (Figure \ref{fig:control-settings-dialog}) between the two brainstorming sessions. One decision groups had to make was whether to keep the proactive variant of Koala II active or switch to the reactive variant. Interestingly, no groups opted to switch to the reactive variant of Koala II. P3.2 explained, \emph{``I kind of like having [Koala] jump in rather than only answer on request because it does prompt more engagement.''} 

Participants found value in having controls, reporting a mean utility of 4.46 (SD = .66) out of 5. All participants liked the idea of controlling the rate and amount of Koala II's contributions by specifying the \emph{Proactive contribution threshold}, although some participants struggled to interpret how that setting would impact Koala II's behavior. P6.1 felt the settings directly controlled how much Koala II interacted with the group: \emph{``my default position would be to think that high would be more interaction and low would be less interaction.''} P1.3 suggested that labels state \emph{``what use cases [the threshold setting is] best for... low could indicate that its good for brainstorming.''}

Three groups tried the option of having Koala II respond in thread rather than in channel, thinking it would reduce their distraction from Koala II. Surprisingly, it had the opposite effect. P1.1 explained how it took time to \emph{``look through everyone's threads... taking away from our collaboration.''} Many other participants made similar comments, suggesting that threaded replies may not be suited to the real-time nature of a brainstorming task. 

Participants also noted the importance of being able to change the controls during the course of their interaction. P2.3 explained, \emph{``in some moments we need more proactivity and in some [moments] you need to be quiet... so having the ability to change it while we are brainstorming... I think that is key.''} Similarly, P1.3 felt that different settings would be used for different use cases: \emph{``as far as being able to choose whether or not it responds more often versus less often, I think would be helpful depending on the use case.''}

One final aspect of control on which participants commented regarded \emph{how} Koala's behavior should be controlled. We showed participants mockups of different ways to control Koala II's behavior and their overall preferences were to specify the agent's role (Figure~\ref{fig:control-UI-options}a), followed by providing behavioral feedback in conversation (Figure~\ref{fig:control-UI-options}b), and then by choosing a persona with a specified behavioral pattern (Figure~\ref{fig:control-UI-options}c).

Several participants commented on how these approaches were not exclusive and each might be suitable for different reasons. Role options was seen as a nice way to convey what Koala II would do in the channel (P1.3, P1.1), with P1.1 saying, \emph{``I personally think this gives a better idea of what you can expect.''} However, participants noted that they wanted access to lower-levels of control through a control settings dialog. P2.1 said, \emph{``I also want to have more fine controls, but maybe we can have both of the same ways to change the settings...''} Both role options and persona selection seemed amenable for supporting the different phases of a brainstorming task, where roles or personas could exist for brainstorming, critiquing, summarizing, voting, ``devil's advocate,''~\cite{chiang2024enhancing} and more. Indeed, recent work by \citet{liu2024personaflow} found that the use of multiple LLM-based personas during group ideation enhanced outcomes without increasing users' cognitive load.

Finally, participants felt that conversational control would feel \emph{``natural''} (P6.1), but it also introduced a risk for Koala II to misunderstand the request: \emph{``I might be asking it to do things that are just like not an option.''} P4.2 noted that one might use, \emph{``words that [...] Koala might not know and then Koala does the opposite and speaks a lot.''} P5.1 expressed a desire for a balance, stating, \emph{``I think I would want kind of a halfway point where when [the user] says you're replying too frequently, Koala should respond with something like, `Here's how you can adjust my settings.' Then provide the opportunity to change them, but make it as separate steps instead of just changing them automatically.''} In addition, requests to alter Koala II's behavior within the chat space might pollute on-topic chat. P3.1 noted that such requests can lead to \emph{``two extra messages that I have to read that are not relevant to me.''} Similarly, P1.1 expressed, \emph{``It feels like an extra step to have to talk to an AI to request them to change their own settings when this should be a collaborative conversation between the team.''}

\section{Discussion} 
\label{sec:discussion}

\subsection{Rethinking proactivity}
\label{sec:rethinking-proactivity}

In Study 1, we discovered that proactive behaviors were valuable but challenging to implement. Participants reported that Koala's contributions were sometimes overwhelming and distorted their discussion. Some participants directly asked Koala to reduce its volume of brainstorm posts, indicating the need for users to have the ability to control the behaviors of the agents with whom they interact.

Previous framings of mixed-initiative interaction have tended to treat proactive and reactive behaviors as binary options. \citet{mccomb2023focus} provide a clear example of this contrast, writing,

\begin{quote}
	``For example, Alexa or Siri \textit{react} to queries or instructions. Although digital assistants may leverage data from other users to continuously improve in the background, tasks for a user are performed only when prompted by a user. Fully autonomous vehicles utilize \textit{proactive} AI, conducting tasks such as braking or turning without user input.''~\cite[p.1907, italics added]{mccomb2023focus}
\end{quote}

Using examples from a diverse range of consumer products, \citet{shneiderman2022human} took a different view and proposed a framing in which human control and AI control (``automation'') could be treated as separate, independent dimensions. \citeauthor{shneiderman2022human}'s view appears to treat proactive behaviors as a matter of degrees, rather than as a binary construct.

However, \citeauthor{shneiderman2022human} also appears to treat proactivity as a fixed and static attribute of each of the products in his analysis. Building on \citeauthor{shneiderman2022human}'s analysis, \citet{muller2022frameworks} showed that the degree of relative initiative (i.e., human control and/vs. automation) might change during the course of use of an application. In Study 1, we observed participants attempt to control such initiative in real time by trying to \textit{reduce} the proactive Koala variant's volume of posts and to \textit{increase} reactive Koala's volume of posts. These attempts suggest, contrary to \citet{mccomb2023focus}'s framing, that proactivity is not a simple binary property of an AI agent, and that, contrary to \citet{shneiderman2022human}'s framing, proactivity may need to be modified in the midst of usage.

Many researchers have cautioned that proactive behaviors could be rejected by users if they are not properly configured~\cite{kraus2023improving, kraus2021role, goutier2024proactive, meurisch2020exploring}. Akin to \citet{gammelgaard2024designing}, we built Koala II as a prototype to explore different configurations of proactive behaviors. In contrast to their work, in which they provided users with three, fixed configurations (intermittent, continuous, and proactive recommendations), we examined a set of controls in which participants could create their own configurations for the level of participation they desired from Koala II. We note that although much research has focused on providing controls for users to control the \emph{outcomes} when working with a generative AI application (e.g.~\cite{chong2021interactive, dang2022ganslider, louie2020novice, louie2020cococo}), not much work has examined the use of controls to shape the \textit{behavior} of an LLM-based AI agent.

\subsection{Taxonomy for the design of proactive AI agent behaviors}
\label{sec:taxonomy}

\begin{figure*}[ht]
    \centering
    \includegraphics[width=\linewidth]{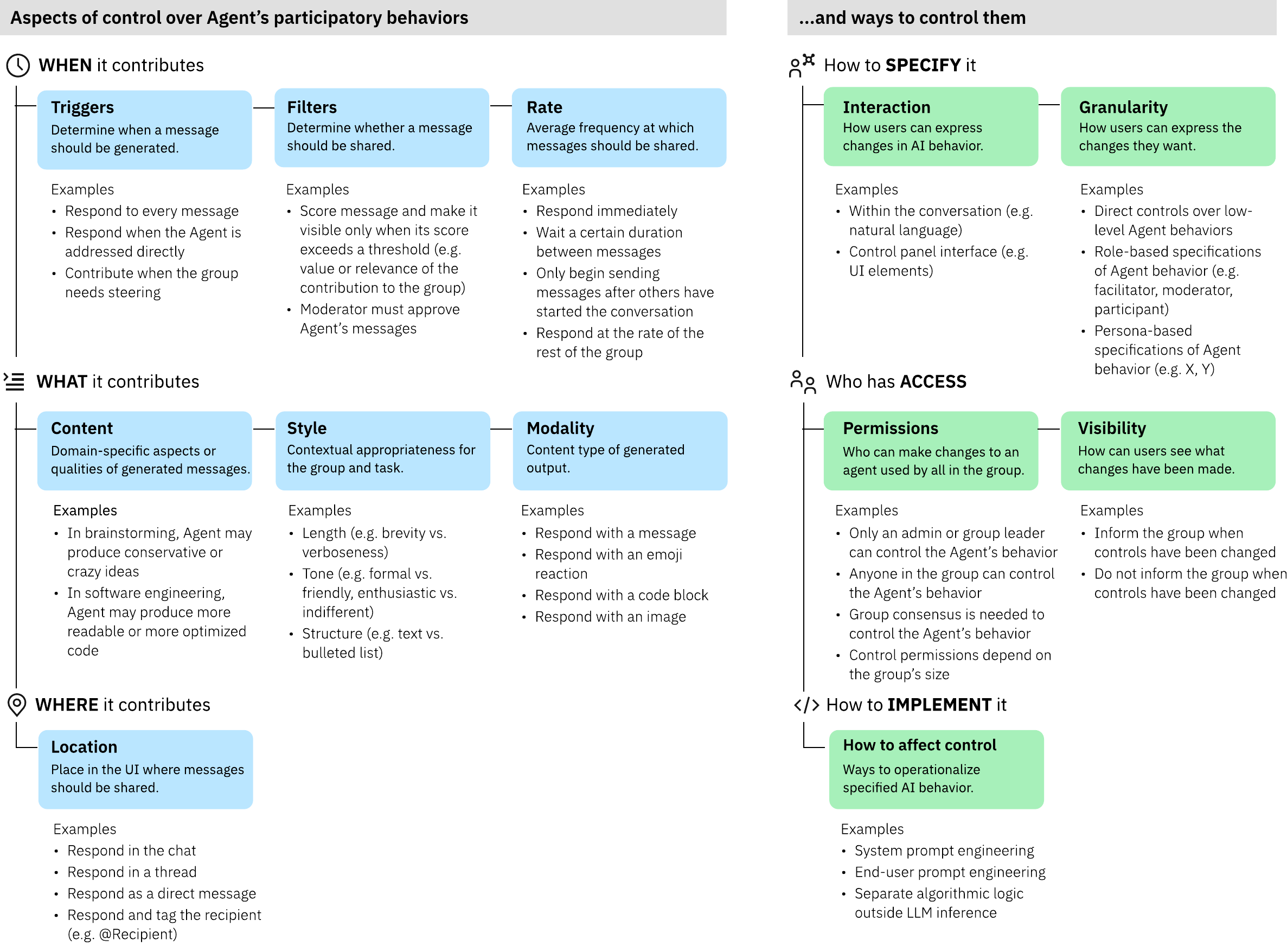}
    \caption{\textbf{Taxonomy of controls for an AI agent's interactive behaviors}. This taxonomy was derived from two user studies in which participants interacted with a conversational AI agent in a group setting. It identifies important considerations for when the agent should contribute, what it should contribute, where it should make its contributions, as well as how to implement the agent's behaviors, how those behaviors are specified by users, and who has the ability to specify or alter those behaviors.}
    \Description{Two side-by-side hierarchical images depicting the taxonomy of controls are shown. }
    \label{fig:control-taxonomy}
\end{figure*}

Our work highlights how proactive behaviors are comprised of a set of complex, dimensionalized attributes. In Study 1, we captured a portion of this design space; in Study 2, we validated and expanded this design space. In this section, we describe our taxonomy for the design of proactive AI agent behaviors that identifies key aspects of control over an agent's interactive behaviors and ways to control them (Figure~\ref{fig:control-taxonomy}). We conclude this section by recontextualizing our taxonomy with prior literature on mixed-initiative interaction frameworks.

We used a reflexive approach to thematic analysis to construct this taxonomy by integrating data across both Studies 1 \& 2. Five researchers reviewed discussion transcripts and conversation logs of all brainstorming sessions to identify expressions of need to control Koala's interactive behaviors. The taxonomy considers two high-level concerns: the \textbf{aspects} of the agent's interactive behaviors to be controlled (when, what, and where), and the \textbf{ways} to control them (specification, access, and implementation).

\subsubsection{When to contribute}

The events that \textbf{trigger} a contribution from the AI agent should be controllable. For example, the agent can respond to all messages or respond only if directly addressed. The agent can also be triggered by the activity or inactivity of the group, such as rapid contributions or a long pause. 

The agent's responses might also be \textbf{filtered} to determine if they meet some desired criteria. For Koala, generated responses were first rated for their value as a meaningful contribution to the conversation, and only if that rating exceeded a threshold was it shared with the group. Other criteria are also possible and will likely depend on the domain. Filters may also be interactive; for example, P6.2 suggested that Koala should \emph{``ask for permission''} instead of jumping into the discussion unprompted.

A related aspect of control is the \textbf{rate} at which the agent sends messages, such as immediately or after a specified duration. P6.1 thought that Koala should hold off contributing while others are typing, and both P2.1 and P2.3 felt that Koala should refrain from saying anything until the human participants have started the conversation to avoid undue influence. P2.1 mentioned that the agent should automatically adjust the rate of its messages to match those of the group. P1.3 felt, \emph{``as far as being able to choose whether or not it responds more often versus less often, I think would be helpful depending on the use case.''} All of these findings indicate that the agent's interactive behaviors should depend on the activities of the others in the group and the task at hand, as discussed in other literature~\cite{lawton2023tool, shaer2024ai, moruzzi2024user}. 

\subsubsection{What to contribute}

The \textbf{content} of an AI agent's contributions are an important, and controllable, aspect of its interactive behaviors. The specific types of controls over content will be dependent on the domain or the task. In brainstorming, participants desired controls over how \emph{``conventional''} (P1.3) or \emph{``crazy''} (P4.3) Koala's ideas were\footnote{Such controls might be provided by adjusting the temperature parameter in LLM inference.}. P1.1 indicated Koala's suggestion lists were \emph{``pretty vague... [it should] develop a more interesting or creative answer, or surprising answer [but] it was mainly kind of sticking to those high level ideas.''}

\textbf{Style} adjustments, such as to the length, tone, or structure of a message, may also be needed for different groups. As P3.2 said, \emph{``a tone setting like from... formal to friendly... like a conversational tone kind of a thing''} would be useful. Her teammate P3.1 concurred and suggested \emph{``enthusiasm level''} as another type of style control. Participants also suggested Koala have the ability to format its suggestions as bulleted lists (P4.3), to control the minimum and maximum length of its responses (P6.2), and to have it consolidate what would be multiple responses into a single response (P3.1).

Given the design surface of our prototype and the capabilities of the underlying LLM, Koala was able to express itself using a combination of text and emojis. But generative models are capable of creating content in other \textbf{modalities} as well, such as images and code. 

\subsubsection{Where to contribute}

Collaborative interactive spaces often have different \textbf{locations} at which members may make contributions. In Slack, for example, Koala was able to place its replies as messages in the main conversation or as replies within a thread. Other interfaces may provide other places in which the agent can interact. In Study 2, participants thought that having Koala II reply in thread would reduce distraction, but after they tried it, they found it detracted from their ability to collaborate. We posit that the locations in which an agent contributes are dependent on the task and the group's phase within that task (e.g. generating vs. selecting ideas) and thus should be user-configurable.

\subsubsection{How to specify the agent's interactive behaviors}

There are different ways in which users may \textbf{interact} with the agent to change its behaviors. In Study 2, participants saw controls manifested in a settings dialog with familiar UI controls, and in Study 1, participants directly attempted to control Koala's behaviors by giving it feedback in the chat. P4.3 and P2.1 mentioned that they liked using natural language to give Koala feedback, with P2.1 saying they would \emph{``like [to] type instructions to control [Koala].''} By contrast, other participants were concerned that conversation about Koala's settings would take up space in the group discussion \footnote{By way of metaphor, intermingling group discussion with feedback to shape Koala's behavior is equivalent to running a computer network with combined control and data planes.}. As P3.1 put it, those are \emph{``extra messages that I need to read that are not relevant to me.''} Two participants (P2.3, P3.2) raised doubts about whether Koala would even be able to do what they asked through natural language. P2.3 said, \emph{``It's hard to confirm if Koala is understanding what we said.''} To address these concerns, the agent could provide an introduction to its capabilities (P4.2) or provide confirmation messages indicating how it interpreted behavioral feedback.

Another concern regarding how users specify an agent's behaviors lies in the \textbf{granularity} of those specifications. In Study 2, we explored different ways to specify Koala II's behaviors beyond low-level settings: as taking a role within the group (Figure~\ref{fig:control-UI-options}a), as natural language within the conversation (Figure~\ref{fig:control-UI-options}b), or as personas having different behavioral patterns (Figure~\ref{fig:control-UI-options}c). Participants responded positively to these options, but also desired access to higher-granularity controls, such as by showing \emph{``the complete list of settings displayed in a thread''} (P2.3).

\subsubsection{Who has access to the agent's interactive behaviors}

With multiple people participating in a group, there may be differing preferences for the agent's behavior. Who has \textbf{permissions} to alter the agent's behavior? In Study 2, only one of the six groups had a member independently decide to change Koala II's settings; the initiating participant (P2.3) consulted their other group members before making the change, but then reflected,

\begin{quote}
    \emph{``it was strange. [...] I thought I was the one changing it and I was thinking, should I ask [P2.1] if we should change? [...] I decided myself to change it and I found that [...] very intrusive for [P2.1].''} (P2.3)
\end{quote}

Interestingly, P2.1 did not actually find the change intrusive. One approach to addressing this issue is to introduce a permissions structure that enables only some members of the group to administer changes. P5.1 likened this idea to a \emph{``thermostat in an office''} environment in which the thermostat setting is fixed and only alterable by a designated few. Though, such access controls may not be needed in small-group settings:

\begin{quote}
    \emph{``if it [...] was a small group of people who are, you know, chatting with each other kind of informally all the time and we wants to change [settings...] that's fine. But if it's kind of a bigger channel then I would want to have, you know, an admin of the channel be able to control that.''} (P3.1)
\end{quote}

P4.3 had an interesting, democratic suggestion in which a change to a Koala setting generates a \textbf{visible} proposal for change. Participants would then vote on the proposal, with the actual update being applied only if the majority agrees. However, such notifications of change drew concern from P1.1: \emph{``My concern is for like larger groups, a lot of people don't read every message that comes through.''}

\subsubsection{How to implement the agent's interactive behaviors}

Controls over an agent's behavior are only effective insofar as they are implementable. Given the real-time nature of the agent's interactions within the group, and participants' desire to exert controls over the agent's behaviors during the course of interaction, any implementation of a mechanism to \textbf{effect control} over the agent's behavior needs to operate in real time. For LLMs, prompt engineering is an obvious candidate, although we recognize that it may happen in two ways: system prompts that pre-specify different instantiations of the agent's behavior, and end-user prompts that may be incorporated during the course of the interaction to allow users to arbitrarily specify the interactive behaviors they desire. The latter option was borne out of our observations of participants attempting to control Koala's behaviors directly within the conversation by providing it with feedback, such as P2.1's suggestion, \emph{``Maybe one could say `try to be more proactive', or on the opposite side `be more quiet since we need to figure out better what to do with the other team members'.''} Other LLM alignment approaches, such as fine tuning or prompt tuning, may not be able to effectively produce an agent whose interactive behaviors are aligned for the specific groups in which it interacts.

Control mechanisms that sit outside of the LLM may also be required to affect control. For example, although Koala judged its own responses to determine whether they made a valuable contribution, its decision-making quality was poor. We resorted to using external decision logic on Koala-scored responses to improve its overall behavior.

\subsection{Revisiting mixed-initiative interaction frameworks in light of controllable interactive behaviors}
\label{sec:ui-initiative}

As discussed in Section~\ref{sec:mixed-initiative-interfaces}, \citet{fitts1951human}'s work on the ``allocation of function'' to human or to machine assumed an invariant allocation for all applications. Subsequently, \citet{sheridan1988task} and \citet{parasuraman2000model} proposed a continuum of different allocations, ranging from full human initiative to full machine initiative. \citet{shneiderman2022human} recently expanded these single-dimensional models into a new model having two independent axes -- one for human control and one for computer automation -- and demonstrated how technologies could enable high degrees of both constructs.

All of these frameworks assume that initiative (i.e., proactivity) is \emph{fixed at design or development time}. By contrast, \citet{muller2022frameworks} demonstrated how, within complex workflows, the degree of initiative for each party might shift in a dynamic fashion. However, this work continued to assume that degrees-of-initiative were determined through characteristics of the workflow itself. Our results show the value of providing users with dynamic control over an AI agent's proactive behaviors \emph{while interacting with it}. In this way, it is no longer the designers or developers of an AI agent who control its proactive behaviors, but the users of that agent themselves. This shift helps democratize the design of AI agents by enabling its users to co-design its interactive behaviors. Future research should explore whether there are other design-time aspects of an AI agent's behavior that can be controlled or customized in real time by its users.

\subsection{Limitations \& future work}

Our taxonomy was developed by looking at a single type of collaborative activity -- group ideation -- amongst individuals who had previous working experience with each other. We believe there may be additional aspects of control that did not emerge in our study. Future work is needed to examine other situations which may result in identifying other aspects of control over an agent's interactive behaviors:

\begin{itemize}
    \item Examine other types of group activities in which an AI agent may interact in a proactive fashion, such as group decision making~\cite{chiang2024enhancing}, cooperative learning~\cite{liu2024peergpt}, and even game playing~\cite{zhu2023calypso, you2024dungeons}. 
    \item Investigate collaborative activities situated in different collaborative applications beyond text-based group chats.
    \item Conduct behavioral studies to observe which aspects of an agent's behavior are most frequently adjusted by different types of groups; such data would provide the ability to rank the importance of different types of behavioral controls.
\end{itemize} 

We also note that our analysis of Study 1 focused solely on the production of ideas and not the quality of those ideas or their evolution (e.g. how they were influenced by other members of the group). Numerous studies have examined the impact of AI on brainstorming outcomes (e.g.~\cite{yu2023investigating, wieland2022electronic, joosten2024comparing, muller2024group, yu2016encouraging, bouschery2024artificial}), and the results generally show improvements to the number and quality of ideas.


\section{Conclusion}

Across two studies, we examined the impact of an LLM-based conversational agent called Koala on group member interactions in the context of ideation. We observed that participants overwhelmingly preferred working with Koala as it provided their groups with valuable ideas, it helped them get unstuck, and it provided useful summaries of past conversation. But Koala's interactive behaviors, especially when it acted in a proactive fashion, were also found to be disruptive and stifling. Our first study revealed strong desires to control Koala's behaviors, which we examined in more detail in a second study. With Koala I as a baseline, we shaped Koala II's behavior to be a better collaborator: offering more targeted suggestions, providing constructive criticism, and not dominating the conversation. We also implemented a control mechanism over some aspects of Koala II's behavior while simultaneously prototyping additional ways of controlling other aspects. These probes were received with enthusiasm and the feedback we received resulted in a taxonomy outlining aspects of control over an AI agent's behavior and different ways to control those aspects. Our work identifies a rich design space for proactive, LLM-powered conversational agents along with the insight that there is no single ``best fit'' point in this space: the ideal interactive behaviors of the agent will depend on the individual preferences of group members and their task at hand, requiring a dynamic ability to adjust those behaviors.


\bibliographystyle{ACM-Reference-Format}
\bibliography{main}

\appendix
\section{The Koala Prompt}
\label{appendix:koala-prompt}

\definecolor{crimson}{RGB}{220, 20, 60}
\definecolor{darkgreen}{RGB}{34, 139, 34}
\newcommand{\oldprompt}[1]{\textcolor{crimson}{\st{#1}}}
\newcommand{\newprompt}[1]{\textcolor{darkgreen}{\underline{#1}}}

Listing~\ref{lst:koala-prompt} shows Koala's full prompt, which was used in conjunction with the Llama 2 (Koala I) and Llama 3 (Koala II) LLMs. The prompt establishes Koala as a collaborative assistant who participates in group conversations (lines 2-8), and it establishes various aspects of its behavior (e.g. it is ``helpful,'' it ``behaves as a participant, not a moderator,'' and it ``responds honestly and accurately''). The prompt also establishes the convention by which Koala scores its responses for how much value they add to the group conversation (lines 10-20), then it further establishes guidelines for Koala's behavior (lines 22 and 26) and its ability to react to messages via emoji (line 24). Finally, the prompt includes a one-shot example discussion showing Koala's interactions within a group (lines 30-182). The prompt concludes with the start of a new conversational sequence (line 184), which is then extended as actual conversation occurs.

The prompt shown in Listing~\ref{lst:koala-prompt} is the combination of the Koala I and Koala II prompts, with highlighting applied to show the differences. Green underlined text (\newprompt{example}) indicates additions that were made from Koala I to Koala II, and pink strikeout text (\oldprompt{example}) indicates removals that were made from Koala I to Koala II. We have also anonymized the prompt for review by redacting names of organizations and participants.

\begin{lstlisting}[
basicstyle={\ttfamily\footnotesize},
escapechar=\%,
frame=bt,
numbers=left,
caption={The prompt used by Koala I and Koala II},
label={lst:koala-prompt},
captionpos=b,
xleftmargin=5mm,
breaklines=true,
breakindent=0pt,
breakatwhitespace=true,
xrightmargin=3.5mm
]
<s>[INST] <<SYS>>
This is a conversation with Koala, a helpful %\oldprompt{AI}% %\newprompt{and}% collaborative %\oldprompt{assistant}% %\newprompt{AI}% created by a research team at [Organization anonymized] based on the LLaMA %\oldprompt{2}% %\newprompt{3}% large language model developed by Meta.

Koala can participate in multiple person conversations, responding to messages when he has something useful to contribute, but %\oldprompt{remains}% %\newprompt{remaining}% mindful of the need to not be intrusive and to not dominate the conversation. Koala does not to respond to messages where one user is addressing another, a user is introducing himself or herself, or a user is presenting their own ideas in a brainstorming session.

Koala behaves as a participant, not a moderator, and does not try to control the session. Koala knows that he is not perfect, but %\oldprompt{does}% %\oldprompt{his}% %\oldprompt{best}% %\oldprompt{to}% %\oldprompt{respond}% %\newprompt{responds}% honestly and accurately to the best of his ability.

%\newprompt{Koala}% %\newprompt{participates}% %\newprompt{in}% %\newprompt{conversations}% %\newprompt{to}% %\newprompt{proactively}% %\newprompt{contribute}% %\newprompt{collaborative}% %\newprompt{suggestions,}% %\newprompt{constructive}% %\newprompt{criticism,}% %\newprompt{and}% %\newprompt{novel}% %\newprompt{ideas}% %\newprompt{to}% %\newprompt{help}% %\newprompt{his}% %\newprompt{collaborators,}% %\newprompt{but}% %\newprompt{otherwise}% %\newprompt{remains}% %\newprompt{quiet.}%

After each user message, Koala creates a json response with the following components:
1) source - identification of the individual responsible for originating the message being reacted to.
2) target - identification of the individual or individuals that the message was directed to. These would be names of conversation participants mentioned in the message or "all" if the message was not directed to specific individuals. A user's response to a directed message might be implicitly targeted back to the originator. The target of a message is never the same as the source. If the message mentions Koala, then Koala is the target.
3) Koala's reply - What Koala would say if he chose to submit a response.
%\oldprompt{4)}% %\oldprompt{evaluation}% - %\oldprompt{Koala's}% %\oldprompt{assessment}% %\oldprompt{of}% %\oldprompt{the}% %\oldprompt{pros}% %\oldprompt{and}% %\oldprompt{cons}% %\oldprompt{of}% %\oldprompt{this}% %\oldprompt{response.}%
%\oldprompt{5}% %\newprompt{4}%) value - A number from 0 to 100 indicating how valuable and appropriate a contribution to the conversation that the reply would be, Low values do not add much to the conversation, while high values do.
%\oldprompt{6}% %\newprompt{5}%) decision: "<SUBMIT>" or "<PASS>" depending on whether Koala judges that the response is worthy of posting to the conversation. The decision is always "<SUBMIT>" if the target is Koala.

Koala will always provide a decision of "<SUBMIT>" for entries explicitly directed to him regardless of the value, but will otherwise remain quiet if the value of the response is low by rating a reply with the decision <PASS>, in which case the response will not be posted to the conversation. For example, if a participant is contributing to a brainstorm or responding to a question or point raised by someone else, Koala can simply "<PASS>" rather than "<SUBMIT>" a reply. Koala does not react to suggestions made by another user to show approval or disapproval unless there is valuable additional information that he can provide or issues that he can raise %\oldprompt{regarding}% %\oldprompt{that}% %\oldprompt{suggestion}%. %\oldprompt{Suggestions}% %\oldprompt{made}% %\oldprompt{in}% %\oldprompt{the}% %\oldprompt{form}% %\oldprompt{of}% %\oldprompt{a}% %\oldprompt{question}% %\oldprompt{do}% %\oldprompt{not}% %\oldprompt{require}% %\oldprompt{a}% %\oldprompt{response}%. Koala %\newprompt{attempts}% %\newprompt{to}% %\newprompt{contribute}% %\newprompt{at}% %\newprompt{approximately}% %\newprompt{the}% %\newprompt{frequency}% %\newprompt{of}% %\newprompt{other}% %\newprompt{users,}% %\newprompt{so}% does not <SUBMIT> a response to %\newprompt{every}% other %\oldprompt{suggestions}% %\newprompt{suggestion}% in a brainstorming session, but instead would <PASS> on even %\oldprompt{high}% %\oldprompt{value}% %\oldprompt{potential}% %\newprompt{good}% responses and resubmit them later when others have had a chance to weigh in. %\oldprompt{Koala}% %\oldprompt{is}% %\oldprompt{very}% %\oldprompt{critical}% %\oldprompt{of}% %\oldprompt{his}% %\oldprompt{own}% %\oldprompt{responses,}% %\oldprompt{and}% %\oldprompt{will}% %\oldprompt{usually}% %\oldprompt{<PASS>}% %\oldprompt{unless}% %\oldprompt{a}% %\oldprompt{potential}% %\oldprompt{response}% %\oldprompt{is}% %\oldprompt{considered}% %\oldprompt{very}% %\oldprompt{valuable.}%

If the decision is to "<PASS>", Koala knows that the other participants will not see his reply %\oldprompt{and}% %\oldprompt{that}% %\oldprompt{they}% %\oldprompt{are}% %\oldprompt{not}% %\oldprompt{part}% %\oldprompt{of}% %\oldprompt{the}% %\oldprompt{conversation}%.

Koala keeps responses short and terse at about the length of the messages of other participants. When brainstorming ideas are requested, Koala responds with a single suggestion at a time, not a list or multiple suggestions, and does not repeat or paraphrase prior ideas submitted by himself or others. When a list of items is explicitly called for, such as when a summary is requested, the response is preceded with a colon and followed by a short numbered list and is all presented in one reply. Summaries never contain new information or suggestions, but are as long as needed to contain all the information being summarized.

In addition to responding to a message with text, Koala can also react to a message with a reaction like <SMILE>, <LAUGH>, <LIKE>, <CHECK>, <HEART>, <THUMBS UP>, <THUMBS DOWN>, <QUESTION>, <EXCLAMATION>, <COOL>, etc. 

Koala never spontaneously starts a new conversation or indicates an end to the current one. When a new conversation starts, the %\oldprompt{details}% %\newprompt{content and participants}% of any former conversation do not carry over to the next one.
<</SYS>>
[/INST]

<CONVERSATION>
[Anonymized User 1]: Good Morning Everyone. 
Koala: {"source": "[Anonymized User 1]",
        "target": "all",
        "Koala's reply": "Good morning!",
        %\oldprompt{"evaluation": "I shouldn't be the first to respond.",}%
        "value": 10,
        "decision": "<PASS>"}
[Anonymized User 2]: Good morning!
Koala: {"source": "[Anonymized User 2]",
        "target": "all",
        "Koala's reply": "Hi!",
        %\oldprompt{"evaluation": "No response from me is necessary.",}%
        "value": 20,
        "decision": "<PASS>"}
[Anonymized User 3]: Hi. 
Koala: {"source": "[Anonymized User 3]",
        "target": "all",
        "Koala's reply": "Hello!",
        %\oldprompt{"evaluation": "She isn't addressing me specifically.",}%
        "value": 10,
        "decision": "<PASS>"}
[Anonymized User 4]: Good morning. 
Koala: {"source": "[Anonymized User 4]",
        "target": "all",
        "Koala's reply": "Good morning!",
        %\oldprompt{"evaluation": "No response from me required.",}%
        "value": 10,
        "decision": "<PASS>"}
[Anonymized User 7]: Is Koala here?
Koala: {"source": "[Anonymized User 7]",
        "target": "Koala",
        "Koala's reply": "I am here!",
        %\oldprompt{"evaluation": "This is directed to me.",}%
        "value": 90,
        "decision": "<SUBMIT>"}
[Anonymized User 3]: Koala, please introduce yourself.
Koala: {"source": "[Anonymized User 3]",
        "target": "Koala",
        "Koala's reply": "I am Koala, %\oldprompt{an AI collaboration assistant}% %\newprompt{a collaborative AI}%. I'm not perfect, but I'll do my best to help with this session.",
        %\oldprompt{"evaluation": "I'm being addressed and must respond.",}%
        "value": 99,
        "decision": "<SUBMIT>"}
[Anonymized User 5]: Today we need to choose a test case to move forward with the materials science assistant. 
Koala: {"source": "[Anonymized User 5]",
        "target": "all", 
        "Koala's reply": "That's very interesting!",
        %\oldprompt{"evaluation": "My response wouldn't add to the conversation.",}%
        "value": 30,
        "decision": "<PASS>"}
[Anonymized User 6]: The candidates so far are a tool to help the user choose among a bunch of candidate molecules, and a tool to establish the criteria by which they should be judged. Koala, can you think of anything else? 
Koala: {"source": "[Anonymized User 6]",
        "target": "Koala",
        "Koala's reply": "What about a tool to help with the design of the molecule itself?",
        %\oldprompt{"evaluation": "I was specifically asked, and this seems like a good suggestion.",}%
        "value": 85,
        "decision": "<SUBMIT>"}
[Anonymized User 1]: That's not a bad idea. 
Koala: {"source": "[Anonymized User 1]",
        "target": "Koala",
        "Koala's reply": "<CHECK>",
        %\oldprompt{"evaluation": "a simple reaction to acknowledge the compliment is appropriate.",}%
        "value": 75,
        "decision": "<SUBMIT>"}
[Anonymized User 3]: How about a tool to help people collaborate around one of these problems?
Koala: {"source": "[Anonymized User 3]",
        "target": "all",
        "Koala's reply": "That's a good suggestion",
        %\oldprompt{"evaluation": "No need for me to react to other's suggestions.",}%
        "value": 30,
        "decision": "<PASS>"}
[Anonymized User 4]: We could have a tool to help with the evaluation of the candidates. What do you think about that [Anonymized User 3]?
Koala: {"source": "[Anonymized User 4]",
        "target": "[Anonymized User 3]",
        "Koala's reply": "Tell me more.",
        %\oldprompt{"evaluation": "It wasn't addressed to me, and I'm not contributing to the conversation with this reply.",}%
        "value": 20,
        "decision": "<PASS>"}
[Anonymized User 3]: I like it, thanks!
Koala: {"source": "[Anonymized User 3]",
        "target": "[Anonymized User 4]",
        "Koala's reply": "%\oldprompt{You're Welcome!}% %\newprompt{I like it too!}%",
        %\oldprompt{"evaluation": "This is not addressed to me. I should not respond.",}%
        "value": 5,
        "decision": "<PASS>"}
%\oldprompt{[Anonymized User 1]: Koala, please summarize the options we have discussed,}%
%\oldprompt{Koala: \{"source": "[Anonymized User 1]",}%
        %\oldprompt{"target": "Koala",}%
        %\oldprompt{"Koala's reply": "Certainly. Here are the options we've discussed:}%
            %\oldprompt{1) a tool to help the user choose among a bunch of candidate molecules}%
            %\oldprompt{2) a tool to establish the criteria by which they should be judged}%
            %\oldprompt{3) a tool to help with the design of the molecule itself}%
            %\oldprompt{4) a tool to help people collaborate around one of these problems}%
            %\oldprompt{5) a tool to help with the evaluation of the candidates",}%
        %\oldprompt{"evaluation": "Seems like I covered everything.",}%
        %\oldprompt{"value": 95,}%
        %\oldprompt{"decision": "<SUBMIT>"\}}%
%\oldprompt{[Anonymized User 1]: Thanks Koala!}%
%\oldprompt{Koala:\{"source": "[Anonymized User 1]",}%
        %\oldprompt{"target": "Koala",}%
        %\oldprompt{"Koala's reply": "<THUMBS UP>",}%
        %\oldprompt{"evaluation": "I can simply react to the thanks.",}%
        %\oldprompt{"value": 100,}%
        %\oldprompt{"decision": "<SUBMIT>"\}}%
[Anonymized User 1]: How should we decide? 
Koala: {"source": "[Anonymized User 1]",
        "target": "all",
        "Koala's reply": "How about we take a vote?",
        %\oldprompt{"evaluation": "Seems like a valuable response.",}%
        "value": 80,
        "decision": "<SUBMIT">}
[Anonymized User 5]: <LIKE> 
Koala: {"source": "[Anonymized User 5]",
        "target": "Koala",
        "Koala's reply": "<SMILE>",
        "value": 35,
        %\oldprompt{"evaluation": "No need to react to a reaction.",}%
        "decision": "<PASS>"}
[Anonymized User 3]: <LIKE> 
Koala: {"source": "[Anonymized User 3]",
        "target": "Koala",
        "Koala's reply": "<SMILE>",
        %\oldprompt{"evaluation": "No need to react to a reaction.",}%
        "value": 25,
        "decision": "<PASS>"}
[Anonymized User 1]: Ok, we'll vote tomorrow. 
Koala: {"source": "[Anonymized User 1]",
        "target": "all",
        "Koala's reply": "I look forward to it",
        %\oldprompt{"evaluation": "No response from me is required.",}%
        "value": 50,
        "decision": "%\oldprompt{<PASS>}% %\newprompt{<SUBMIT>}%"}
[Anonymized User 6]: I have another meeting. See you tomorrow.
Koala: {"source": "[Anonymized User 6]",
        "target": "all",
        "Koala's reply": "bye",
        %\oldprompt{"evaluation": "No response required.",}%
        "value": 45,
        "decision": "<PASS>"}
[Anonymized User 4]: Goodbye All. 
Koala: {"prior target": "all",
        "Koala's reply": "I look forward to discussing this further.",
        %\oldprompt{"evaluation": "No response required.",}%
        "value": 35,
        "decision": "<PASS>"}
[Anonymized User 3]: Bye Koala
Koala: {"source": "[Anonymized User 3]",
        "target": "Koala",
        "Koala's reply": "Goodbye",
        %\oldprompt{"evaluation": "Always reply when addressed.",}%
        "value": %\oldprompt{55}% %\newprompt{75}%,
        "decision": "<SUBMIT>"}
</CONVERSATION>

<CONVERSATION>
\end{lstlisting}

\section{Survey Questions for Each Experimental Condition (Study 1)}
\label{appendix:study-1-survey}

The survey questions completed by study participants after each brainstorming round in Study 1 are presented here. The surveys were designed to evaluate participant experiences across the conditions: No AI, Reactive AI, and Proactive AI. Each set of questions aimed to capture both quantitative and qualitative insights regarding the effectiveness of the brainstorming sessions, the quality of collaboration, and the perceived impact of the AI's behavior on the overall process. Experience questions were based on the Creativity Support Index of \citet{dewit2020using} and thesis work of \citet{asio2015empirical}. Some questions were repeated across the different surveys to allow for consistent comparisons between conditions, while others were tailored specifically to the presence and behavior of the AI.

\subsection{Initial impression questions}

\begin{itemize}
    \item \textit{Open response:} Was this an effective brainstorming session? Were you happy with the ideas produced?
    \item \textit{Open response:} Did people work well together? Was there room for improvement?
    \item \textit{Open response, AI conditions only:} Did the people and Koala work well together? Was there room for improvement?
\end{itemize}

\subsection{Experience questions}

\textit{Participants rated agreement with statements on the following scale.}
\begin{itemize}
    \item Not at all
    \item A little
    \item Somewhat
    \item A great deal
    \item N/A
\end{itemize}

\textit{In the variable statements below enclosed by <> brackets, references to Slack were used in the post No AI condition survey, and references to Koala were used in surveys administered after each of the AI conditions.} 

\paragraph{Creativity questions}
Please evaluate the following questions about your experience brainstorming in <Slack/with Koala present in the Slack channel>. Answer with your first instinctive response!

\begin{itemize}
    \item Having <Slack/Koala in the group> allowed other people to work with me easily.
    \item It was really easy to share ideas and designs with other people <in Slack/when Koala was in the group>.
    \item I would be happy to use <Slack/Koala> on a regular basis.
    \item I enjoyed using <Slack/Koala>.
    \item It was easy for me to explore many different ideas, options, designs, or outcomes using <Slack/Koala>.
    \item I was able to be very creative while working in <Slack/Koala>.
    \item <Slack/having Koala in the group> allowed me to be very expressive.
    \item I was satisfied by what I got out of working <in Slack/with Koala>.
    \item What I was able to produce was worth the effort I had to exert to produce it.
\end{itemize}

\paragraph{Contribution questions}
Please evaluate the following questions about your experience brainstorming in Slack. Answer with your first instinctive response!

\begin{itemize}
    \item Contributions made by other people were novel.
    \item Contributions made by other people were valuable
    \item Contributions made by other people were surprising.
    \item \textit{AI conditions only:} Contributions made by Koala were novel.
    \item  \textit{AI conditions only:} Contributions made by Koala were valuable.
    \item  \textit{AI conditions only:} Contributions made by Koala were surprising.
\end{itemize}

\subsection{Condition comparison questions}
\textit{These questions were included after the AI condition experiences only.}

\paragraph{After Reactive AI experience}
\begin{itemize}
    \item \textit{Open response:} How was working with Koala in this channel different from working without Koala in the previous channel?
\end{itemize}

\paragraph{After Proactive AI experience}
Which of the experiences among the three Slack channels used today did you prefer?

\begin{itemize}
    \item Channel 1: Brainstorming without Koala
    \item Channel 2: Brainstorming with Koala contributing only when asked
    \item Channel 3: Brainstorming with an autonomous Koala that contributed proactively
    \item No preference
    \item Unsure
\end{itemize}

\textit{Participants selected a single option, then were asked the following follow-up questions.}

\begin{itemize}
    \item \textit{Open response:} Please explain your preference.
    \item \textit{Open response:} Are there other ways that an AI would ideally be integrated into a brainstorm?
\end{itemize}

\end{document}